\pdfoutput=1
\RequirePackage{lineno}
\newif\ifpreprint%
\preprintfalse%
\ifpreprint%
	\documentclass[preprint,english,aps,prb,floatfix,amssymb,superscriptaddress,a4paper]{revtex4}
\else%
	\documentclass[twocolumn,english,aps,prb,floatfix,amssymb,superscriptaddress,a4paper]{revtex4}
\fi%
\usepackage{amsmath}
\usepackage{siunitx}
\usepackage{dsfont}
\usepackage{color}
\usepackage{amssymb}
\usepackage{graphicx}
\usepackage{braket}
\usepackage{units}
\usepackage{chemformula}
\usepackage{amssymb}
\usepackage{mathtools}
\usepackage{physics}
\usepackage{braket}
\usepackage{bibunits}

\usepackage{hyperref}
\hypersetup{
  colorlinks = true,
  hypertexnames=true
}
\newcommand{\ssm}{\scriptscriptstyle\rm}

\renewcommand{\theta}{\vartheta}
\renewcommand{\phi}{\varphi}

\newcommand{\RN}[1]{%
  \textup{\uppercase\expandafter{\romannumeral#1}}%
}

\begin{document}
\ifpreprint%
	\linenumbers%
\fi%

\title{Axial-field-induced chiral channels in an acoustic Weyl system}

\author{Valerio Peri}
\affiliation{Institute for Theoretical Physics, ETH Zurich, 8093 Z\"urich, Switzerland}
\author{Marc Serra-Garcia}
\affiliation{Institute for Theoretical Physics, ETH Zurich, 8093 Z\"urich, Switzerland}
\author{Roni Ilan}
\affiliation{Raymond and Beverly Sackler School of Physics and Astronomy, Tel-Aviv University, Tel-Aviv 69978, Israel}
\author{Sebastian D. Huber}
\affiliation{Institute for Theoretical Physics, ETH Zurich, 8093 Z\"urich, Switzerland}

\let\oldaddcontentsline\addcontentsline
\renewcommand{\addcontentsline}[3]{}
\begin{bibunit}[naturemag]
\date{\today}

\maketitle

{\bf 
Condensed-matter and other engineered systems, such as cold atoms,\cite{Roy18} photonic,\cite{Lu15} or phononic metamaterials,\cite{Li17} have proven to be versatile platforms for the observation of low-energy counterparts of elementary particles from relativistic field theories. These include the celebrated Majorana modes,\cite{Mourik12} as well as Dirac\cite{Borisenko14, Liu14a} and Weyl fermions.\cite{Weng15, Xu15,Lv15} An intriguing feature of the Weyl equation\cite{Weyl29} is the chiral symmetry, where the two chiral sectors have an independent gauge freedom. While this freedom leads to a quantum anomaly,\cite{Adler69, Bell69,Bertlmann00, Landsteiner16, Gooth17} there is no corresponding axial background field coupling differently to opposite chiralities in quantum electrodynamics. Here, we provide the experimental characterization of the effect of such an axial field in an acoustic metamaterial. We implement the axial field through an inhomogeneous potential\cite{Pikulin16} and observe the induced chiral Landau levels. From the metamaterials perspective these chiral channels open the possibility for the observation of non-local Weyl orbits\cite{Potter14} and might enable unidirectional bulk transport in a time-reversal invariant system.\cite{Huber16}
}

Three-dimensional Weyl semimetals are characterized by a touching of two non-degenerate Bloch bands. Around the touching point, the low-energy physics can be described by an equation akin to Weyl's equation \cite{Weyl29} for massless relativistic particles 
\begin{linenomath}
\begin{equation}
        \label{eqn:weyl}
H=\sum_{\alpha,\beta=x,y,z} v_{\alpha\beta} (\delta k_\alpha+k_\alpha^{\ssm WP}) \sigma_\beta,
\end{equation}
\end{linenomath}
where ${\bf k}^{\ssm WP}$ is the location of the touching point and $v_{\alpha\beta}$ denotes the velocity tensor. The Pauli-matrices $\sigma_\beta$ encode some pseudo-spin degree of freedom reflecting the two involved bands. A key property of Weyl systems is their chirality $s={\rm sign}[\det(v_{\alpha\beta})]=\pm 1$, which controls if (pseudo-) spin and momentum are aligned or anti-aligned. Moreover, the eigenstates of Eq.~(\ref{eqn:weyl}) define a monopole source for the Berry-curvature, which in turn forces such WPs to appear in pairs of opposite chirality.\cite{Nielsen83}

Much of the recent interest in Weyl systems \cite{Borisenko14, Liu14a, Weng15, Bernevig15, Xu15, Yang15a, Lv15} arises from their magneto-transport properties.\cite{Nielsen83, Potter14, Moll16} The application of a magnetic field ${\bf B}$ leads to Landau levels that are dispersing along the field direction and have zero group velocity perpendicular to it.\cite{Nielsen83} In particular, the dispersion of the zeroth Landau levels 
\begin{linenomath}
\begin{equation}
        \label{eqn:dispersion}
\omega({\bf k}) =  {\rm sign}(B_{\parallel})s v_{\parallel}  k_{\parallel}
\end{equation}
\end{linenomath}
depends on the chirality $s$ of the WPs and on the projection of the WP velocity onto the field direction $v_{\parallel}$.\cite{Landsteiner16} This makes Weyl systems the momentum-space bulk analog of the quantum Hall effect:\cite{Klitzing80} Unidirectional channels are separated in momentum space rather than in real space. Moreover, these chiral channels live in the three dimensional bulk as opposed to on the edge of a two-dimensional sample. Such unidirectional channels might harbor interesting physical effects or technological promises\cite{Huber16} also for classical systems such as acoustic \cite{Xiao15b, Li17, He18} or electromagnetic metamaterials.\cite{Lu15} However, as neither phonons nor photons carry an electromagnetic charge $e$, it remained unclear if such chiral Landau levels can be observed in neutral metamaterials.

\begin{figure}[tbh]
\includegraphics{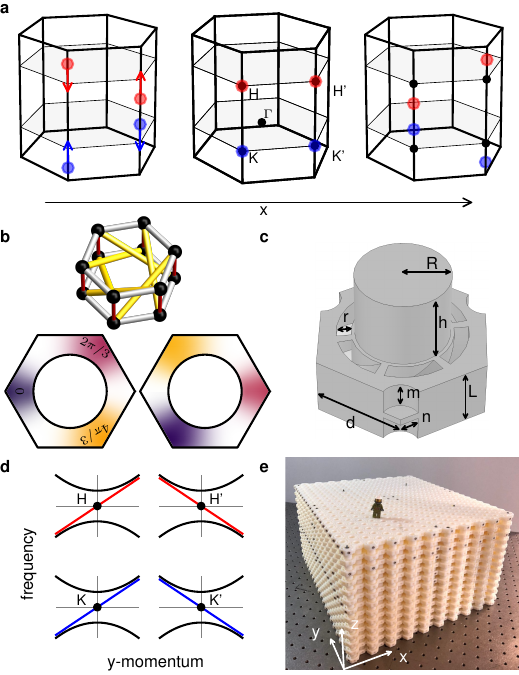}
\caption{
{\bf Inhomogeneous Weyl point separation and axial magnetic fields.} {\bf a,} Hexagonal Brillouin zone with the high-symmetry points $\Gamma$,$K$,$K'$,$H$, and $H'$. The blue (red) dots indicate schematically Weyl points of chirality $s=+1$ ($-1$) whose location in momentum space is shifted as a function of the spatial coordinate $x$ due to the presence of an axial magnetic field. {\bf b,} Tight binding model and the local acoustic orbitals in the structure shown in {\bf c}. The density plot indicates the pressure field in the mid-plane of the the cavity at $h/2$. The colors mark the phases of the local modes. {\bf c,} Unit-cell of the hexagonal acoustic crystal. The parameters are explained in the main text. {\bf d,} Chiral Landau levels emerging at the high-symmetry points due to the axial magnetic field. The color again indicates the chirality of the Weyl points. {\bf e,} Photo of the acoustic crystal.
} 
\label{fig:concept}
\end{figure}

Looking at Eq.~(\ref{eqn:weyl}), we observe that the effect of a magnetic field can be viewed as a space-dependent shift of the WP locations ${\bf k}^{\ssm WP}(x)$. This is most easily seen in the Landau gauge. For a magnetic field ${\bf B}=B\hat{\bf e}_y$, the momentum  
\begin{linenomath}
\begin{equation}
k_z\rightarrow k_z+eBx
\end{equation}
\end{linenomath}
is shifted in space as a function of $x$ as a result of minimally coupling the corresponding gauge field. Taking this viewpoint of inhomogeneous WP positions, it has recently been realized that the application of other space dependent perturbations can lead to effects alike the ones induced by a magnetic field both for graphene\cite{Vozmediano10} as well as for Weyl semimetals\cite{Liu13, Cortjio15, Grushin16, Pikulin16, Sumiyoshi16, Abbaszadeh17} For example, in reaction to an inhomogeneous uniaxial strain, the locations of the WPs are moved in space.\cite{Levy10, Grushin16, Pikulin16} The main difference to a real magnetic field is the chirality dependence of the shift
\begin{linenomath}
\begin{equation}
        \label{eqn:axial}
        k_z \rightarrow k_z+sB_{5}x,
\end{equation}
\end{linenomath}
where $B_{5}$ is the y-component of an axial magnetic field\cite{Liu13, Landsteiner16} and moves WPs of opposite chiralities $s$ in opposite directions, cf. Fig.~\ref{fig:concept}a. Interestingly, the axial nature of the field, together with the chirality-dependence of Eq.~(\ref{eqn:dispersion}) leads to co-propagating chiral channels in the presence of an axial magnetic field. Here, we set out to measure the effect of an axial (background) field in an acoustic system, i.e. in pressure waves in air, where the conceptual idea of an inhomogeneous WP location leads to a set of chiral Landau levels.

In a series of recent papers\cite{Xiao15b, Li17} a tight-binding model for an acoustic Weyl system has been proposed\cite{Xiao15b} and implemented.\cite{Li17} The starting point of the model of Ref.~\onlinecite{Xiao15b} are honeycomb layers with an in-plane hopping $t_n$ shown in gray in Fig.~\ref{fig:concept}b. The stacked honeycomb layers are coupled via small direct hoppings $t_d$ (red) and large chiral hoppings $t_c$ (yellow) which strongly break inversion symmetry. The resulting spectrum is characterized by the minimal number of four WPs for time reversal invariant systems.\cite{Liu13} Moreover, the strong inversion symmetry breaking leads to a maximal separation of the WPs which lie in the vicinity of the high symmetry points $K$, $K'$, $H$, $H'$ shown in Fig.~\ref{fig:concept}a (see Methods). 

We can now induce an axial field via an effective space-dependent sub-lattice potential $\propto B_{5}x\sigma_z$.\cite{Roy18} As the WPs with different chirality $s$ differ by the sign of $v_{zz}$ in Eq.~(\ref{eqn:weyl}), such a global term creates an axial field $\propto sB_{5}x\sigma_z$ in the low-energy theory (Methods). In other words, the WPs shift as shown in Fig.~\ref{fig:concept}a. 
\begin{figure*}[tbh]
\includegraphics{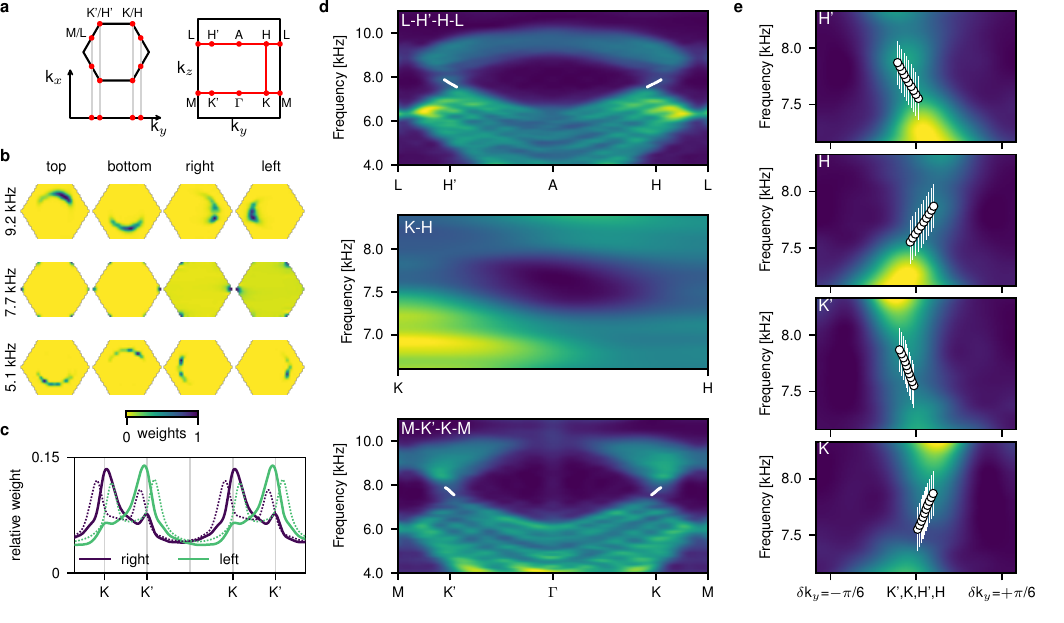}
\caption{
{\bf Observation of chiral Landau levels.} {\bf a,} High-symmetry lines in the Brillouin zone and projection of the $x$-axis onto $k_y,k_z$. {\bf b,} Fourier transform to the first Brillouin zone at $k_z=0$ ($k_z=\pi/a_z$ shows similar behavior which is not shown.) of the acoustic response at different frequencies for different speaker positions (top, bottom, right, or left, see text). The concentration of the response at $7.7\,{\rm kHz}$ around the $K$ and $K'$ points shows that physics at these frequencies is governed by the band-touchings at the corners of the Brillouin zone. The excitation sensitivity of the $K$ and $K'$ at $7.7\,{\rm kHz}$ strongly indicates chiral channels, see text. At lower (higher) frequencies, the response is dominated by a spherically symmetric spectrum.  {\bf c,} Quantitative analysis of the above density plots for $7.7\,{\rm kHz}$ (solid lines) and $7.0\,{\rm kHz}$ (dashed lines), showing that at $7.0\,{\rm kHz}$ the response is already dominated by the $K$ and $K'$ points. At $7.7\,{\rm kHz}$, the chiral Landau levels lead to an imbalance between $K$ and $K'$ as a function of excitation location. {\bf d,} Measured spectrum along the high-symmetry lines indicated in {\bf a}. Both for the top and bottom panel one can observe the touching of two bands around $7.7\,{\rm kHz}$. Overlaid is the analysis presented in {\bf e}. The middle panel shows a gap opening along the $K$-$H$ lines. {\bf e,} Dispersion along the chiral Landau levels at the four high-symmetry points in the Brillouin zone. The momentum-frequency relation was obtained via fitting the phase evolution of the measured frequency response (see text). The horizontal error-bars, reflecting the error in fitting, are smaller than the symbol size. The vertical error bars indicate our frequency uncertainty given the dissipation in the acoustic field.
} 
\label{fig:main}
\end{figure*}

To take the step from the tight-binding model to a concrete acoustic structure we need to find a geometry, such that dissipative acoustic waves behave like a Weyl system in some frequency range. We start from the unit-cell shown in Fig.~\ref{fig:concept}c. The pillars of radius $R$ and height $h$ create a hexagonal layer. The in-plane modes at the $K$ and $K'$-points are shown in Fig.~\ref{fig:concept}b. We see that the low-frequency physics around these points is governed by modes localized at the corners of the unit cell, i.e., on a honeycomb lattice with nearest neighbor distance $d$. The width $r$ of the ``ventilator holes'' controls the strength of the interlayer coupling, while the turning angle $\vartheta$ of the ventilators determines the ratio $t_d/t_c$. Finally, we design a gradient along the $x$-direction by introducing holes above one of the two sub-lattices with varying radius $n$ and fixed depth $m$ (see Methods for details).

In summary, for our setup [see Fig.~\ref{fig:concept}e] we expect four WPs, where the pair $K$ and $H$ sees an axial field $sB_{5}$. As displayed in Fig.~\ref{fig:concept}a and Eq.~(\ref{eqn:dispersion}), the two opposite chiralities at $K$ and $H$ shift in opposite directions under the influence of the axial gauge field. Since $B_5$ does not break time-reversal symmetry, the time-reversed partners $K'$ and $H'$ feel an opposite field $-sB_{5}$, such that $K$ ($H$) and $K'$ ($H'$) shift in opposite directions. Together with Eq.~(\ref{eqn:dispersion}) this leads to the chiral Landau levels depicted in Fig.~\ref{fig:concept}d.

To characterize the properties of the three-dimensional sample we measure the spectral response of the acoustic field. We excite sound waves at different locations on the surface of the system. The response is then measured via a sub-wavelength microphone on the inside of the system. Using a lock-in measurement we measure phase and amplitude information of the acoustic field (Methods). This amounts to the measurement of the Greens function $G({\bf r}_i, {\bf r}_j, \omega )= \langle \psi^*_{i}(\omega)\psi_{j}(\omega)\rangle$, where $\psi_i(\omega)$ is the acoustic field at site ${\bf r}_i$ and frequency $\omega$. By taking the spatial Fourier-transform of this signal, we obtain the spectral response shown in Fig.~\ref{fig:main}. 

Owing to the gauge choice for the field $B_{5}$, the momentum in $x$-direction is not well defined. We therefore show the projection of the spectra onto $k_y$ and $k_z$ along the high-symmetry lines shown in Fig.~\ref{fig:main}a. The top and bottom panels of Fig~\ref{fig:main}d display the touching points of two bands at the $H'$, $H$ and $K'$ and $K$ points, respectively. We further analyze the nature of these touching points through their associated surface physics below. The middle panel shows the evolution of the band-structure from $K$ to $H$, indicating a gap opening in $k_z$ direction assuring that we indeed couple different layers via the ``ventilators''.

\begin{figure}[tbh]
\includegraphics{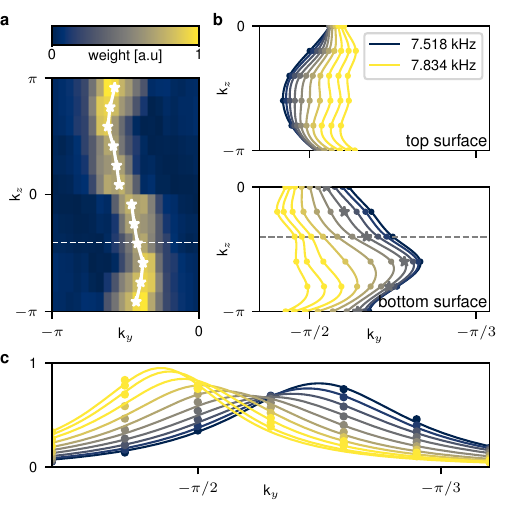}
\caption{
{\bf Characterisation of surface Fermi arcs.} {\bf a,} Surface Fourier-transform of the measured Fermi arc on the $+\hat {\bf x}$ surface. The white curved lines show the location of the maximal response. This is obtained fitting the response at fixed frequency and $k_z$ with a Lorentzian in order to find the $k_y$ with maximal response as shown in {\bf c} for $k_z$ marked with a dashed line. {\bf b,} Evolution of the white curved lines in {\bf a} as a function of frequency $\omega$ on the two surfaces $\pm \hat {\bf x}$ (top, bottom). The opposite sign of the group velocity $v_y=\partial_{k_y}\omega$ indicate the chiral nature of the surface modes winding around the sample according to the $k_z$-layer Chern number.
} 
\label{fig:surface}
\end{figure}

Around the frequency of the WPs, most of the acoustic response is concentrated around the $K$, $K'$, $H$, and $H'$ points. In Fig.~\ref{fig:main}b, we present the response for three selected frequencies in the $k_z=0$ plane. The whole 2D Brillouin zone is presented for clarity, although $k_x$ is not a conserved quantity. The four columns show the response when excited from the 2D surfaces with surface normal $\pm \hat {\bf x}$ (top, bottom), from the surface $\pm \hat {\bf y}$ (right, left), respectively. Three observations can be made: (i) Only those modes which have a group-velocity that allows energy to be transported into the bulk are excited. (ii) Around $7.7\,{\rm kHz}$, only modes around the $K$ and $K'$ are involved. (iii) When excited from the left (right), only the $K$ ($K'$) are excited. However, no such $K$ -- $K'$ selectivity occurs for the top/bottom excitation. These three observations together essentially prove that we deal with a system with chiral channels in $k_y$-direction around this frequency. This sensitivity on the excitation point is further quantified in Fig.~\ref{fig:main}c, where we show the integrated weights of Fig.~\ref{fig:main}b for all $x$ along $k_y$ at the $k_z=0$ plane.

The open boundary conditions on the $ \pm \hat{\bf y}$ (right/left) surfaces allow for a further analysis of the chiral channels. Radiation into free space essentially allows for any mode with arbitrary wave number to be supported inside our sample. In particular, no finite size quantization $k_\alpha=2\pi m/L_\alpha$, with $\alpha=y,z$ and $m\in \mathbb Z$ occurs. We use the fact that the eigenstates are Bloch waves 
\begin{linenomath}
\begin{equation}
        \label{eqn:bloch}
        \psi_i({\bf k})=e^{i{\bf k}\cdot {\bf r}_i} 
        \begin{pmatrix}
                u_{{\bf k}} \\
                v_{{\bf k}}
        \end{pmatrix},
\end{equation}
\end{linenomath}
where $u_{\bf k}$ and $v_{\bf k}$ are the sub-lattice weights. Using the above structure, we can fit the phase evolution from unit-cell to unit-cell to obtain $k_y(\omega)$, see Methods for details. This analysis is only possible owing to the fact that with the excitation-point selectivity shown in Fig.~\ref{fig:main}b/c, we can ensure that in the chiral channel region we fit the phase evolution of only {\em one mode}. The resulting dispersion curves $\omega(k_y)$ are shown in Fig.~\ref{fig:main}e. The clearly visible four chiral channels in accordance with the expectations in Fig.~\ref{fig:concept}d are the main result of this study. 

By measuring the group velocities and the gap opened around the WPs we can determine the magnitude of the axial field. Going through the detailed analysis, we find the in-plane ``velocities'' $v_n=18.01\pm 1.22\,{\rm kHz}^2$, out-of-plane velocity $v_z=7.68\pm0.78\,{\rm kHz}^2$ and an axial field corresponding to a flux per plaquette of $\Phi_{B_5} \approx 0.08 / 2\pi$ in units where the flux quantum is given by $\Phi_0=h/e=1$, see the Supplementary Materials for a detailed explanation of the units and procedures. 

Given the strong inhomogeneity represented by the gradient in the hole diameters, it is worthwhile to validate that we still deal with the physics of acoustic Weyl points. The existence of surface channels ending at the approximate location of the bulk Weyl points (the celebrate Fermi arcs for electronic systems) is such an indication of Weyl physics.

The experimentally measured Fermi arcs are shown in Fig.~\ref{fig:surface}a. From the shown Green's function at a given frequency on the left, we can extract $\omega(k_y,k_z)$. Owing to our resolution in $k_z$ and the dissipation-broadened features, we cannot determine the end-points of this Fermi arc. However, we can observe the evolution of $\omega(k_y,k_z)$ and hence determine the group velocity on the two opposing surfaces. In Fig~\ref{fig:surface}b the surfaces $+ \hat{\bf x}$ (called top surface before) and $-\hat{\bf x}$ (called bottom surface) show indeed opposite group velocities, in line with the expectation of edges states induced by a Chern number. The upper half of the Brillouin zone is not shown and determined by time-reversal symmetry.

Further studies of the effects of axial gauge field in transport phenomena require wave packets experiments. The losses of our system, mainly due to the surfaces, prevent us from performing such experiments. There are currently other acoustic platforms\cite{Ni18} that meet the quality factor requirements to perform such experiments. In the light of recent discoveries of loss-induced exceptional rings in Weyl semimetals,\cite{Xu17z} it is interesting to note that here, dissipation mainly leads to spectral broadening. We extend on that point in the Supplementary Information.

 By directly observing chiral Landau levels in a Weyl system we have shown that axial fields rooted in the theory of high-energy physics can be implemented and observed in condensed matter systems and implemented for the first time a gauge field in an acoustic 3D system.\cite{Yang17, Wen18} Many of the theoretically predicted phenomena, such as the chiral magnetic effect,\cite{Fukushima08} the chiral vortical effect,\cite{Landsteiner16} strain induced quantum oscillations\cite{Liu17} and many more seem now to be reachable in systems of classical metamaterials,\cite{Lee17, Lu15, Li17, Fruchart18b} in cold-atoms, or in low-temperature electronic systems. 
 

\bigskip
\noindent
{\bf Acknowledgements} We acknowledge insightful discussions with Dmitry Pikulin and Ady Stern. We are grateful for the financial support from the Swiss National Science Foundation, the NCCR QSIT, and the ERC project TopMechMat.

\smallskip
\noindent
{\bf Author contributions} SDH, RI and VP performed to the theoretical part of this work. MSG and VP conducted the experiments. All authors contributed to the writing of the manuscript.

\smallskip
\noindent
{\bf Data availablility} The data that support the plots within this paper and other findings of this study are available from the corresponding author upon reasonable request.

\medskip
\noindent
\clearpage
{\bf \large Methods}\medskip

{\bf Details of the sample design and fabrication.} The gradient in $x$-direction in the geometry induces changes to both the real-part of the eigenvalues as well as to the dissipation. The quantitative determination of dissipative effects in the full crystal is computationally hard owing to the high aspect ratio of the boundary layers and the viscoelasticity of the structure. Such a calculation is beyond the scope of this work. For the sample design we only rely on two key ingredients: (i) A strong inversion-symmetry breaking through the twisted interlayer coupling that induces Weyl physics. (ii) A breaking of the the sixfold rotational symmetry via the gradient in $x$-direction giving rise to an axial field.

We adjust the parameters of the unit cells via full-wave simulations in order to fix the WPs at a constant frequency, to avoid any tilt in the Weyl cones,\cite{Soluyanov15} and to minimize spatial variations in the velocities of the linear dispersion. The final sample is then constructed by assembling unit-cells with the pre-determined parameters to obtain the sought after gradient term leading to $\propto B_{5}x\sigma_z$.

The sample shown in Fig.~\ref{fig:concept}e is printed on a Stratasys Connex Objet500 with PolyJet technology by Stratasys. The printed material is VeroWhitePlus. The $300\,\mu{\rm m}$ resolution in the $xy$-plane and a resolution of $30\,\mu{\rm m}$ in the $z$-direction assure a fine surface finish for good hard-wall boundary condition for the acoustic field. At least in the frequency/wavelength regime we are interested in. Each layer was printed in four parts and later assembled and stacked. 

The full sample consists of $L_x\times L_y\times L_z=20\times20\times12$ unit cells. The fixed dimensions of the sample are given by $\vartheta=2.7\,{\rm rad}$, $d=13\,{\rm mm}$, $a_z=18.5\,{\rm mm}$, $R+r=9\,{\rm mm}$. The depth $m=L/2-0.5\,{\rm mm}$ is fixed to leave the separating wall between layers of constant thickness of $1\,{\rm mm}$. The remaining parameters are optimized as described in the main text and are varied between $0\,{\rm mm}\leq n \leq 3.0\,{\rm mm}$, $6.1\,{\rm mm} \leq R\leq 7.6\,{\rm mm}$, and $8.0\,{\rm mm} \leq h \leq 9.3\,{\rm mm}$. The detailed profile of these parameters are presented in Table~\ref{tab:geoParam}. 
\begin{table}[h!]
\centering
\begin{tabular}{l | c  c  c }
   & $h$  ($\rm mm$)&  $R$ ($\rm mm$)&  $n$ ($\rm mm$) \\
	\hline
	 UC 1  & 8.5 & 6.5 & 2.6 \\
   UC 2  & 8.0 & 7.3 & 1.9 \\
   UC 3  & 8.3 & 6.9 & 2.1 \\
   UC 4  & 8.5 & 6.8 & 2.0 \\
   UC 5  & 8.7 & 6.8 & 1.9 \\
   UC 6  & 8.3 & 7.4 & 1.3 \\
   UC 7  & 8.3 & 7.5 & 1.1 \\
   UC 8  & 8.9 & 6.9 & 1.2 \\
   UC 9  & 9.0 & 7.2 & 0.4 \\
   UC 10 & 9.3 & 6.8 & 0.0 \\
   UC 11 & 8.9 & 7.4 & 0.4 \\
   UC 12 & 8.7 & 7.2 & 1.0 \\
   UC 13 & 8.2 & 7.6 & 1.1 \\
   UC 14 & 8.4 & 7.3 & 1.4 \\
   UC 15 & 8.7 & 6.8 & 1.9 \\
   UC 16 & 8.5 & 6.8 & 2.0 \\
   UC 17 & 8.3 & 6.9 & 2.1 \\
   UC 18 & 8.0 & 7.3 & 1.9 \\
   UC 19 & 8.5 & 6.5 & 2.6 \\
   UC 20 & 8.7 & 6.1 & 3.0 \\
	\hline
	\end{tabular}
  \caption{\label{tab:geoParam}Values of the geometrical parameters for each unit cell (UC) of the acoustic crystal. }
\end{table}	
%

The sample is terminated with open surfaces on the $\pm \hat{\bf z}$ (ventilators) and $\pm \hat{\bf y}$ (armchair) edges. These open boundary conditions allow for the phase fitting mentioned in the main text. Along the $\pm \hat{{\bf x}}$ surface we close the sample with hard walls. In their absence the outermost mode localized on the corner of the unit cell in Fig.~\ref{fig:concept}b is largely detuned. This leads effectively to a bearded edge, where spurious surface states appear around the $\Gamma$ and $A$ point, see Supplementary Information for details. To avoid these states, we close the sample on the  $\pm \hat{{\bf x}}$  surfaces. On each $\pm \hat{{\bf x}}$ surface, there is a small hole in the hard boundary that allows for the insertion of speaker and the excitation from that surface.

\noindent
{\bf Tight-binding model.} The dynamics governed by the tight-binding model of Fig.~\ref{fig:concept}b can be written as 
\begin{linenomath}
\begin{equation}
        \label{eqn:tight}
\partial_{t}^2
\begin{pmatrix}
u_{\bf k} \\ v_{\bf k} 
\end{pmatrix}
=\underbrace{
\begin{pmatrix}
\gamma({\bf k})+\alpha({\bf k})& \beta({\bf k})\\
\beta^*({\bf k})& \gamma({\bf k})-\alpha({\bf k})
\end{pmatrix}}_{\mathcal D({\bf k})}
\begin{pmatrix}
u_{\bf k} \\ v_{\bf k} 
\end{pmatrix},
\end{equation}
\end{linenomath}
with 
\begin{linenomath}
\begin{align*}
\gamma({\bf k})&=2\cos(k_z a_z)\bigg[t_d+t_c\cos(k_ya_n)\\&\phantom{=}+2t_c\cos(\sqrt{3}k_xa_n/2)\cos(k_ya_n/2) \bigg]+\epsilon_0^2 ,\\
\beta({\bf k})&=t_n\bigg[e^{- i\sqrt{3}k_x a_n/3}+2\cos(k_ya_n/2) e^{i\sqrt{3}k_x a_n/6}\bigg],\\
\alpha({\bf k})&=2t_c \sin(k_z a_z)\bigg[\sin(k_y a_n)\\
&\phantom{=}-2\cos(\sqrt{3}k_x a_n/2)\sin(k_y a_n/2)\bigg].
\end{align*}
\end{linenomath}
Here, $u_{\bf k}$ and $v_{\bf k}$ describe the amplitudes of the Bloch waves of Eq.~(\ref{eqn:bloch}). The lattice constants are related to the sample geometry through $a_n=\sqrt{3}d$ and $a_z=L+h$. Note, however, that we only match the low-frequency physics around the WPs of model (\ref{eqn:tight}) to the corresponding low-frequency physics of the acoustic structure. In particular, we do not intend to match the full lattice model at all lattice momenta.

The model (\ref{eqn:tight}) features band touchings whenever $\alpha({\bf k})=\beta({\bf k})=0$.\cite{Xiao15b} The function $\beta({\bf k})$ describes a simple honeycomb layer and hence has gap closings at the 
 $K_{\ssm 2d}=(0,-4\pi/3a_n)$ and $K_{\ssm 2d}'=(2\pi/\sqrt{3}a_n,-2\pi/3a_n)$ points. A straight-forward analysis of $\alpha({\bf k})$ shows that the full model has WPs at the $K=(0,-4\pi/3a_n,0)$, $K'=(2\pi/\sqrt{3}a_n, -2\pi/3a_n, 0)$, $H=(0,-4\pi/3a_n, \pi/a_z)$, and $H'=(2\pi/\sqrt{3}a_n, -2\pi/3a_n, \pi/a_z)$ points, respectively, cf. Fig.~\ref{fig:concept}a. 

The term $\gamma({\bf k})$ in Eq.~(\ref{eqn:tight}) introduces a frequency shift or tilt of the conical dispersion. To avoid that, we need $\gamma({\bf k})=0$ for ${\bf k}=K,\,K',\,H,\,H'$. A straight-forward analysis shows how this happens whenever $2t_d=3t_c$.  
 
The low-frequency physics around the Weyl points at frequency $\epsilon_0$ is then fully described by the velocity tensors
\begin{linenomath}
\begin{equation}
        {\mathcal D}({\bf k})=\epsilon_0^2+\sum_{\alpha\beta}v_{\alpha\beta}k_{\beta}\sigma_\alpha,
\end{equation}
\end{linenomath}
which are given by
\begin{linenomath}
\begin{align*}
        v_{\alpha\beta}^{K/K'}&=
        \begin{pmatrix}
             0 & \sqrt{3}t_na_n/2 & 0 \\
             \pm \sqrt{3}t_na_n/2 & 0 &0 \\
             0&0&\pm3\sqrt{3}t_c a_z  
        \end{pmatrix},\\
        v_{\alpha\beta}^{H/H'}&=
        \begin{pmatrix}
             0 & \sqrt{3}t_na_n/2 & 0 \\
             \pm \sqrt{3}t_na_n/2 & 0 &0 \\
             0&0&\mp3\sqrt{3}t_c a_z  
        \end{pmatrix}.
\end{align*}
\end{linenomath}
Going from $K$ ($H$) to $K'$ ($H'$) two rows change sign. When comparing $K$ and $H$, on the other hand, only one row differs in sign. This explains the distribution of chiralities (red and blue) in Fig.~\ref{fig:concept}a.

From the above velocity tensors we see that only $k_z$ couples to the $\sigma_z$ matrix. If we now want to shift the WP in $k_z$ direction as $k_z\rightarrow k_z+sB_5x$, we need to couple a space dependent sub-lattice potential of the form 
\begin{linenomath}
\begin{equation}
 V(x)=3 \sqrt{3}t_c a_z B_5 x\sigma_z\;.
\end{equation}
\end{linenomath}
It is crucial to note that the axial nature of the field arises from the chirality dependent pre-factor of $v_{zz}$. In other words, the above potential $V(x)$ acquires the axial nature only in the low-energy theory. In the Supplementary Material we show how we fit the experimental results to obtain the effective parameters of the tight-binding model.
 
\noindent
{\bf Chirality and Berry curvature.} For completeness, we present the derivation of the Berry-monopole represented by a WP. A WP is a conical touching of two bands, hence its general Hamiltonian can be written as
\begin{linenomath}
\begin{equation}
\label{eqn:hh}
H=\sum_\alpha d_\alpha(\mathbf{k})\sigma_\alpha\;,
\end{equation}
\end{linenomath}
where $\mathbf{d}(\mathbf{k})$ is a vector linear in $\mathbf{k}$. in polar coordinates $\mathbf{d}= |\mathbf{d}|\,(\sin{\theta}\cos{\phi}\,,\, \sin{\theta}\sin{\phi} \,,\, \cos{\theta})$. The eigenvalues of the Hamiltonian~(\ref{eqn:hh}) are $\epsilon(\mathbf{k})=\pm|\mathbf{d}(\mathbf{k})|$. The eigenvectors are
\begin{linenomath}
\begin{equation}
\ket{-}=\begin{pmatrix}\sin{\left(\frac{\theta}{2}\right)}e^{i\phi} \\ -\cos{\left(\frac{\theta}{2}\right)}\end{pmatrix} \qquad \ket{+}=\begin{pmatrix} \cos{\left(\frac{\theta}{2}\right)}e^{i\phi} \\ \sin{\left(\frac{\theta}{2}\right)}\end{pmatrix}\;.
\end{equation}
\end{linenomath}

To characterize the WP the lower band is relevant. Neglecting $\mathcal{A}_{|d|}=i\langle -|{\partial_{|d|}}|-\rangle$, the remaining components of $\boldsymbol{\mathcal{A}}$ are
\begin{linenomath}
\begin{align}
\mathcal{A}_{\theta}&=i\langle -|\partial_{\theta}|-\rangle=0, \\ \mathcal{A}_{\phi}&=i\langle -| \partial_{\phi}|-\rangle=\sin^2{\left(\frac{\theta}{2}\right)}.
\end{align}
\end{linenomath}
From this, the Berry curvature $\boldsymbol{\mathcal{F}}=\nabla \wedge  \boldsymbol{\mathcal{A}}$ follows:
\begin{linenomath}
\begin{equation}
\mathcal{F}_{|d|}=\frac{\sin{\theta}}{2}\;.
\end{equation}
\end{linenomath}
To get back to the original coordinates, this result needs to be multiplied by the Jacobian of the coordinate transformation. If $\mathbf{d}(\mathbf{k})=\mathbf{k}$, the Jacobian is the one of a spherical coordinate transformation ($1/\sin{\theta}|\mathbf{k}|^2$) and the Berry curvature is
\begin{linenomath}
\begin{equation}
\label{eqn:partialB}
\boldsymbol{\mathcal{F}}=\frac{\mathbf{k}}{2|\mathbf{k}|^3}\;.
\end{equation}
\end{linenomath}
Integrated over a shell around WP, this amounts to a flux of $2\pi$, i.e., the WP corresponds to a monopole charge. However, so far we have considered the case of isotropic WPs. 

In the most general case the Weyl Hamiltonian is given by Eq.~(\ref{eqn:weyl}). In this case $\mathbf{d}(\mathbf{k})\neq\mathbf{k}$. However, there is still a linear relation between $\mathbf{d}$ and $\mathbf{k}$. If the tensor $v_{ \alpha\beta}$ was diagonal it would amount to a simple rescaling. In the most general case, this transformation amounts to a permutation of the rows of the Jacobian and a rescaling. The parity of this permutation is captured by the determinant of the velocity tensor. Therefore, the Berry curvature for the generic Weyl point described by Eq.~(\ref{eqn:weyl}), is
\begin{linenomath}
\begin{equation}
\boldsymbol{\mathcal{F}}=
{\rm sign}[\det(v_{\alpha\beta})]\,
\frac{\mathbf{k}}{2|\mathbf{k}|^3}=s\, 
\frac{\mathbf{k}}{2|\mathbf{k}|^3}.
\end{equation}
\end{linenomath}
In other words, the chirality of the WP defines the charge of the Berry monopole.

\noindent
{\bf Measurement and signal analysis.} The acoustic signals are generated with speakers SR-32453-000 from Knowles. The pressure fields are measured via a sub-wavelength microphone FG-23629-P16 from Knowles with a diameter of $2.6\,{\rm mm}$ which is mounted on a $2\,{\rm mm}$ steel rod to scan the inside of the acoustic crystal. We always measure 200 frequency points between $4$ and $11\,{\rm kHz}$ to obtain phase and amplitude information using a lock-in amplifier. In all measurements, different background levels are attributed to slightly different speaker positions.

In unit-cell coordinates $(i_x,i_y,i_z)$ running from $i_x,i_y=0,\dots,20$ and $i_z=0,\dots,12$ we excited at $(0,10,6)$ [called bottom in the main text], $(20,10,6)$ [top], $(10,0,6)$ [left], and $(10,20,6)$ [right]. The crystal is scanned in a grid of $19\times20
\times12\times2$ points corresponding to all the accessible sub-lattice sites of a set of 12 stacked layers of each $20\times20$ honeycomb unit-cells. The surfaces with normal $\pm {\bf x}$ are terminated with hard walls in order to prevent the appearance of spurious surface modes (see Supplementary Information for details). Note that a potential finite size quantization of the wave number $k_x$ is irrelevant here, as the gradient in $x$-direction of the sample geometry renders it ill-defined anyway. The surfaces with normal $\pm {\bf y}$ and $\pm {\bf z}$ are kept open for reasons explained below.

The zig-zag terminations along the surfaces $\pm\hat{\bf x}$ limit the number of points that can be measured at the surfaces. The data displayed in Fig.~\ref{fig:main} are based on pure bulk measurement. The first two unit cells closest to the $\pm\hat{\bf x}$ and the closest unit cells to $\pm\hat{\bf z}$ and $\pm\hat{\bf y}$ surfaces have not been take into consideration in these analysis to avoid spurious surface effects. On the other hand, the data of Fig~\ref{fig:surface} are based on surface measurements only, i.e., data taken on the first two unit-cells. Finally, for the density plots, the discrete spatial Fourier transforms are displayed with the Lanczos interpolation method for visual clarity, except in Fig~\ref{fig:surface}b, where the discrete nature of the measurements is relevant. Note, that the chiral channel phase fitting in Fig~\ref{fig:main}e is not based on an interpolation of the Fourier-transform data.

\noindent
{\bf Chiral channel phase fitting.}
The eigenstates of the system are Bloch waves
\begin{linenomath}
\begin{equation}
        \psi_i({\bf k})=\begin{pmatrix}
                \psi_i^A({\bf k})\\
                \psi_i^B({\bf k})
        \end{pmatrix}=e^{i{\bf k}\cdot {\bf r}_i} 
        \begin{pmatrix}
                u_{{\bf k}} \\
                v_{{\bf k}}
        \end{pmatrix},
\end{equation}
\end{linenomath}
where $u_{\bf k}$ and $v_{\bf k}$ are the sub-lattice weights. As prescribed by Bloch's theorem, the phase of this eigenstates evolves linearly as a function of position. Hence, the momentum $k_y$ of the eigenstate can be extracted by fitting the phase evolution of the signal as a function of $y$. 

We take the discrete Fourier transform along $x$ and $z$ of the pressure field measured on one of the two sub-lattices in each unit-cell to obtain $\psi^{A/B}(k_x,k_z;y)$. By fitting a linear function to the evolution of the phase of the Bloch wave
\begin{linenomath}
\begin{equation}
\arg[\psi^{A/B}(k_x^{\alpha},k_z^{\alpha};y)]\approx k_y  y
\end{equation}
\end{linenomath}
we can extract the momentum $k_y$ of the chiral channel at each frequency for the different WPs at $(k_x^\alpha,k_z^\alpha)$. Note that selecting both $k_x$ and $k_z$ does not limit the analysis. In fact, both $K$ ($H$) and $K'$ ($H'$) points exist with the same momentum in the $x$ and $z$ direction but different along $y$. To prevent the participation of other bulk modes, we limit the frequency rage to the size of the gap opened around the WPs, cf. Supplementary Material. 

For the channels at $K$ and $H$ the data obtained with excitation on the $-\hat{\bf y}$ surface were used, while for the channel at $K'$ and $H'$ the ones from surface $\hat{\bf y}$. This is essential since to apply the phase fitting procedure only one mode should be excited at the time. The horizontal error-bars in Fig~\ref{fig:main}e, obtained from the fitting procedure, are smaller than the symbol size. The vertical error bars indicate our frequency uncertainty given the dissipation in the acoustic field $\Delta \omega \approx 200\,{\rm Hz}$. Moreover, we limit the channel analysis to frequencies within the gap determined in the supplementary information. Note, that the fact that we obtain a smooth set of $k_y$-values from this fitting which has a resolution much below the finite size quantization of $\Delta k_y\approx 2\pi/L_y d$, supports our claim that we have essentially open boundary conditions.


\end{bibunit}

\let\addcontentsline\oldaddcontentsline

\renewcommand{\thetable}{S\arabic{table}}
\newcommand{\ph}{\phantom\dagger}
\renewcommand{\thefigure}{S\arabic{figure}}
\renewcommand{\theequation}{S\arabic{equation}}
\onecolumngrid
\pagebreak
\setcounter{page}{1}
\thispagestyle{empty}
\begin{center}
	\textbf{\large Supplemental Material: Axial-field-induced chiral channels in an acoustic Weyl system}\\[.2cm]
	
	  Valerio Peri,$^{1}$ Marc Serra-Garcia,$^{1}$ Roni Ilan,$^{2}$ and Sebastian D. Huber$^{1}$ \\[.1cm]
	  {\itshape ${}^1$Institute for Theoretical Physics, ETH Zurich, 8093 Z\"urich, Switzerland\\
	  ${}^2$Raymond and Beverly Sackler School of Physics and Astronomy, Tel-Aviv University, Tel-Aviv 69978, Israel\\}
	(Dated: \today)\\[1cm]
	\end{center}
	\appendix
	\renewcommand{\thesection}{\Roman{section}}
	\begin{bibunit}[naturemag]
	\section{Effective tight-binding model}

The sample design was motivated by the low-energy physics around the Weyl point (WP) of the tight-binding model introduced in Ref.~\onlinecite{Xiao15b} and illustrated in Fig.~1b of the main text. In order to extract the strength of the axial field $B_5$ from the measurement, we fit the physics around the WPs to the low-energy theory given in Eq. (7) of the main text. For completeness, we state the full model here
\begin{linenomath}
\begin{equation}
        \label{eqn:tight}
\partial_{t}^2
\begin{pmatrix}
u_{\bf k} \\ v_{\bf k} 
\end{pmatrix}
=\underbrace{
\begin{pmatrix}
\gamma({\bf k})+\alpha({\bf k})& \beta({\bf k})\\
\beta^*({\bf k})& \gamma({\bf k})-\alpha({\bf k})
\end{pmatrix}}_{\mathcal D({\bf k})}
\begin{pmatrix}
u_{\bf k} \\ v_{\bf k} 
\end{pmatrix},
\end{equation}
\end{linenomath}
with
\begin{linenomath}
\begin{align*}
\gamma({\bf k})&=2\cos(k_z a_z)\bigg[t_d+t_c\cos(k_ya_n)\\&\phantom{=}+2t_c\cos(\sqrt{3}k_xa_n/2)\cos(k_ya_n/2) \bigg]+\epsilon_0^2 ,\\
\beta({\bf k})&=t_n\bigg[e^{- i\sqrt{3}k_x a_n/3}+2\cos(k_ya_n/2) e^{i\sqrt{3}k_x a_n/6}\bigg],\\
\alpha({\bf k})&=2t_c \sin(k_z a_z)\bigg[\sin(k_y a_n)\\
&\phantom{=}-2\cos(\sqrt{3}k_x a_n/2)\sin(k_y a_n/2)\bigg].
\end{align*}
\end{linenomath}
We can then write the low-frequency expansion around the WPs as
\begin{linenomath}
\begin{equation}
\ddot {\bf u}  = -D {\bf u} =-\left[ \epsilon_0^2 {\bf 1} \pm  
v_n \tilde k_x \sigma_y + v_n \tilde k_y \sigma_x \pm v_z (\tilde k_z\pm \tilde B_5 \tilde x) \sigma_z \right]{\bf u}.
\end{equation}
\end{linenomath}
The velocities are given by 
\begin{linenomath}
\begin{align}
        v_n&= \frac{\sqrt{3}}{2}t_n, \\
        v_z&= 3 \sqrt{3} t_c,
\end{align}
\end{linenomath}
and we defined $\tilde k_\alpha=k_\alpha a_n$ for $\alpha=x,y$, $\tilde k_z=k_z a_z$, and $\tilde x = x/a_n$. In other words we measure the momenta free of length scales, i.e. [$\tilde k_x$]=[$\tilde k_y$]=[$\tilde k_z$]=1 such that the $K$-points are at $(\pm 2\pi/\sqrt{3},\pm 2\pi/3,0)$ or $(0,\pm 4\pi/3,0)$ and similarly for the $H$, $M$, $A$, and $\Gamma$ points. Moreover $\tilde B_5 = 2\pi B_5 a_n a_z  /\Phi_0$, where $\Phi_0=h/e$ denotes the magnetic flux quantum. This immediately implies that
\begin{linenomath}
\begin{align}
    [\epsilon_0] &= {\rm Hz},\\
    [v_n]=[v_z]  &= {\rm Hz^2},\\
    [\tilde B_5] &=1.
\end{align}
\end{linenomath}

Standard manipulations lead to the eigenvalues $\lambda$ of $D$ in the form of 
\begin{linenomath}
\begin{equation}
   \lambda_\pm = \epsilon_0^2 \pm \sqrt{2mv_nv_z\tilde B_5+v_y^2\tilde k_y^2} \qquad m\in \mathbb N.  
\end{equation}
\end{linenomath}
The eigenfrequencies of the acoustic system are given by the square roots of $\lambda$. For $m=0$ we obtain the chiral Landau level
\begin{linenomath}
\begin{equation}
\omega_{\rm LL} = \sqrt{\epsilon_0^2 \pm |v_n  k_y |},
\end{equation}
\end{linenomath}
whereas for $m=1$ the spectrum is gapped at the $K$, $K'$, $H$, and $H'$ points
\begin{linenomath}
\begin{equation}
\omega_{\rm g}^{\pm} = \sqrt{\epsilon_0^2 \pm \sqrt{2 v_n v_z \tilde B_5}}.
\end{equation}
\end{linenomath}
Finally, close to the WPs, we obtain 
\begin{linenomath}
\begin{align}
 \omega_{\rm LL}&\approx \frac{v_n}{2 \epsilon_0} k_y+\epsilon_0, \\
 \Delta = \omega_{\rm g}^+-\omega_{\rm g}^- &\approx \frac{\sqrt{2v_nv_z \tilde B_5}}{\epsilon_0}.
\end{align}
\end{linenomath}
It is now evident that we need to extract $\epsilon_0$, $\Delta$, $v_n$, and $v_z$ to measure our axial field strength $\tilde B_5$.
\begin{figure*}[tb]
        \begin{center}
                \includegraphics{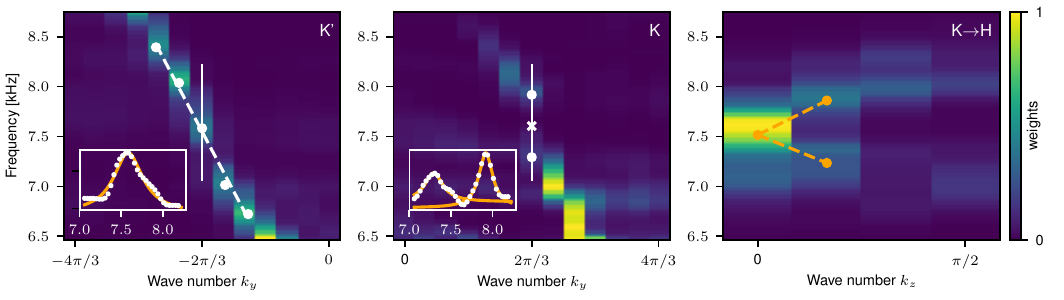}
        \end{center}
        \caption{{\bf Extraction of tight-binding parameters} The left panel shows the chiral Landau level excited from the surface with normal $+\hat {\bf y}$. The fact that we observe only one peak at $k_y=-2\pi/3$ (inset) illustrates the linearly dispersion Landau level. From the slope we extract the velocity $v_n$ (see text). The middle panel shows the same for $k_y$ in the vicinity of $K$, where the same excitation has no overlap with the chiral Landau level with positive group velocity. Consequently, we find two peaks at $k_y=2\pi/3$ corresponding to the first gapped Landau levels. The right panel shows how one can extract the velocity $v_z$ (see text). Note that the panels above show the response when only excited from {\em one} side as opposed to Fig.~2 of the main text where different excitation directions are overlaid in panels d\&e.}
        \label{fig:supmatextract}
\end{figure*}

Using the directional dependence of our excitation scheme we can extract all needed parameters. The left and middle panel of Fig.~\ref{fig:supmatextract} show the Fourier transform of the acoustic field when excited from the surface with normal $+\hat{\bf y}$ evaluated at $k_x$ and $k_z$ corresponding to the $K$ and $K'$ points. The presence of the axial field is clearly visible by the fact that for $K'$, where the chiral Landau level has negative group velocity, the spectrum is given by a straight line with one peak per momentum. For $K$, on the other hand, the chiral Landau level has positive group velocity and can hence not be excited from the surface with normal $+\hat{\bf y}$. Indeed, at $\tilde k_y=2\pi/3$, corresponding to the $K$-point, we find two peaks indicating the gapped Landau levels with $m=1$. Using $\Delta$ and $\epsilon_0$ extracted from the middle panel we can determine $v_n$ from the slope in the left panel. 

The velocity $v_z$ can be obtained from the response along the line $K\to H$. In principle, one could expect to find a flat Landau level in $\tilde k_z$ direction. However, given the small gap $\Delta$ and the small number of unit cells, the extent of the flat level is very limited and at $\tilde k_z = 2\pi/12 $ (the first discrete $\tilde k_z$ value for our sample) we are already in the linear regime of the original WP. We substantiate this below in Sec.~\ref{sec:flat}.

The resulting parameters are given by 
\begin{linenomath}
\begin{align}
        \Delta		&=  \phantom{1}0.60  \pm  0.11\, {\rm kHz},      \\
        \epsilon_0	&=  \phantom{1}7.71  \pm  0.06\, {\rm kHz},      \\
        v_n		&= 18.01  \pm  1.22\, {\rm kHz}^2,  \\
        v_z		&=  \phantom{1}7.68  \pm  0.78\, {\rm kHz}^2,  \\
        t_n		&= 20.80  \pm  1.41\, {\rm kHz}^2,  \\
        t_c		&=  \phantom{1}1.48  \pm  0.15\, {\rm kHz}^2,  \\
        t_n/t_c		&= 14.07  \pm  2.38,          \\
        \tilde B_5      &=  \phantom{1}0.08  \pm  0.03 \approx 0.01 \times 2\pi.          
\end{align}
\end{linenomath}
The last expression for $\tilde B_5$ is relevant as $\tilde B_5=2\pi$ would amount to a full flux quantum per plaquette. The uncertainties in the parameters arise from the statistical errors given that we fit all points $K$, $K'$, $H$, and $H'$ with the respective excitation directions. It may be helpful to note, that the above parameters results in group velocities $\tilde v_n$, $\tilde v_z$ in meters per second through
\begin{align}
    \tilde v_n = \frac{v_n a_n}{2\epsilon_0} &\approx 26.3\, {\rm m/s}, \\
    \tilde v_z = \frac{v_z a_z}{2\epsilon_0} &\approx \phantom{2}9.2\, {\rm m/s}.
\end{align}

We now calculate the strength of the magnetic induction for standard electronic Weyl semimetals corresponding to our $\tilde B_5$. The field strength is given by 
\begin{linenomath}
\begin{equation}
    B  = \frac{\tilde B_5 \Phi_0}{2\pi A},
\end{equation}
\end{linenomath}
where $\Phi_0=h/e\approx4.134\times 10^{-15}\,{\rm Wb}$, and $A$ is the area of the unit cell. Using the lattice constants ($3.43$\,\AA,$11.6$\,\AA), ($12.6$\,\AA,$25.4$\,\AA) for ${\rm TaAs}$ and ${\rm Cd}_3{\rm As}_2$, respectively, we find
\begin{linenomath}
\begin{align}
        {\rm TaAs}:\quad B&\approx 126\, {\rm T}\;\; ({\rm short}),\;\; 
        B\approx 427\, {\rm T}\;\; ({\rm long}),\\
        {\rm Cd}_3{\rm As}_2:\quad B&\approx 16\, {\rm T}\phantom{2}\;\; ({\rm short}),\;\; 
        B\approx 32\, {\rm T}\;\; ({\rm long}),
\end{align}
\end{linenomath}
depending if the field is applied perpendicular to the short or long crystalline axis.

\begin{figure*}[tb]
        \begin{center}
                \includegraphics{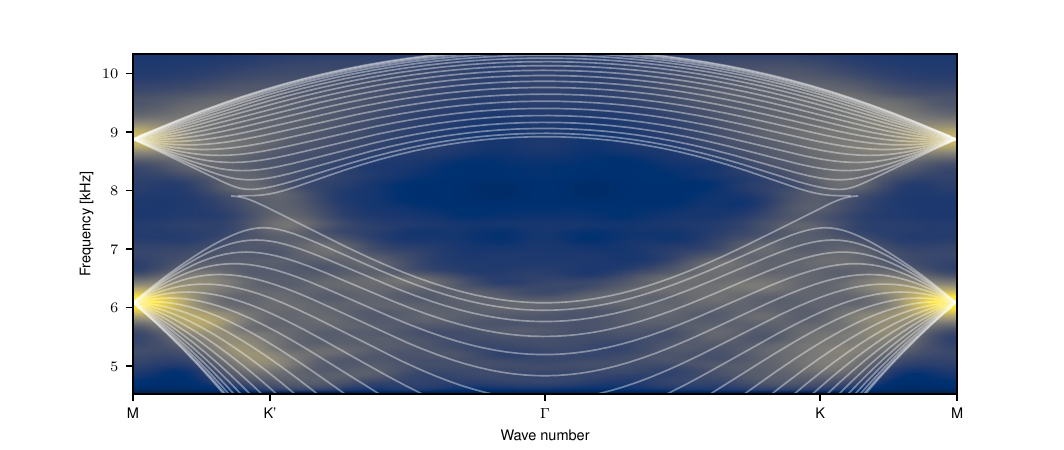}
        \end{center}
        \caption{Overlay of the effective tight-binding model fitted to the data. Note that we only show theoretical lines, where the weight of the mode is concentrated in the bulk, rather than on the surface of the system. See text for the details of the fitting procedure.}
        \label{fig:supmatoverlay}
\end{figure*}

The tight-binding model fitted to the low-frequency physics around the WPs can be overlaid with the data as shown in Fig.~\ref{fig:supmatoverlay}. Note that there is no reason the expect that the tight-binding model fits the regions far away from the WP in a quantitative way.

\section{Boundary terminations on $\pm\hat{\bf x}$ surfaces}
\noindent
In the chiral channel analysis the measurement points closer to the boundaries are not taken into account in order to be sure to consider only bulk modes. However, the termination of the sample is relevant to determine the location of the Fermi arcs in momentum space. Studies of the tight binding model presented in Methods, shown in Fig.~\ref{fig:suppBoundary}a, illustrate the location along $k_y$ of the surface modes displayed in red. The simulated model has the same number of unit cells along $x$ as the metamaterial and also the coupling parameters and the gauge field have been chosen to fit the acoustic crystal low-energy. In the case of zig-zag termination, Fig.~\ref{fig:suppTBField}, the Fermi arcs are present for $k_y \in [-4\pi/3\,,-2\pi/3]$ and $k_y \in [2\pi/3\,,4\pi/3]$. On the other hand, for a bearded termination of Fig.~\ref{fig:suppBoundary}a the arcs are locate at $k_y\in [-2\pi/3\,,2\pi/3]$. Closing our sample on $\pm\hat{\bf x}$, we implement a zigzag termination on the $\pm\hat{\bf x}$ surfaces as proved by the location of the Fermi arcs shown in main text.
\ifpreprint%
\else%
\begin{figure}[tb]
\includegraphics{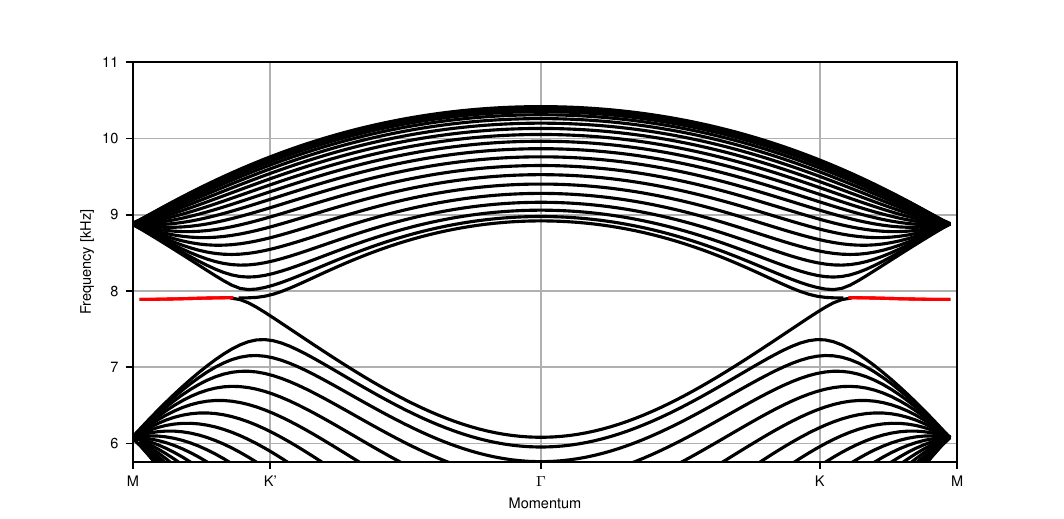}
\caption{{\bf Chiral channels in tight binding model.} The band structure of the tight binding model with an axial gauge field is presented. The system is finite along $x$ with $L_x=20$ and periodic along the other directions, with zigzag boundaries on $\pm\hat{\bf x}$. The tight binding parameters are the same as the ones used in Fig.~\ref{fig:supmatoverlay}. The red modes are localized on the $\pm\hat{\bf x}$ surfaces.}
\label{fig:suppTBField}
\end{figure}
\fi%

\ifpreprint%
\else%
\begin{figure}[tb]
\includegraphics{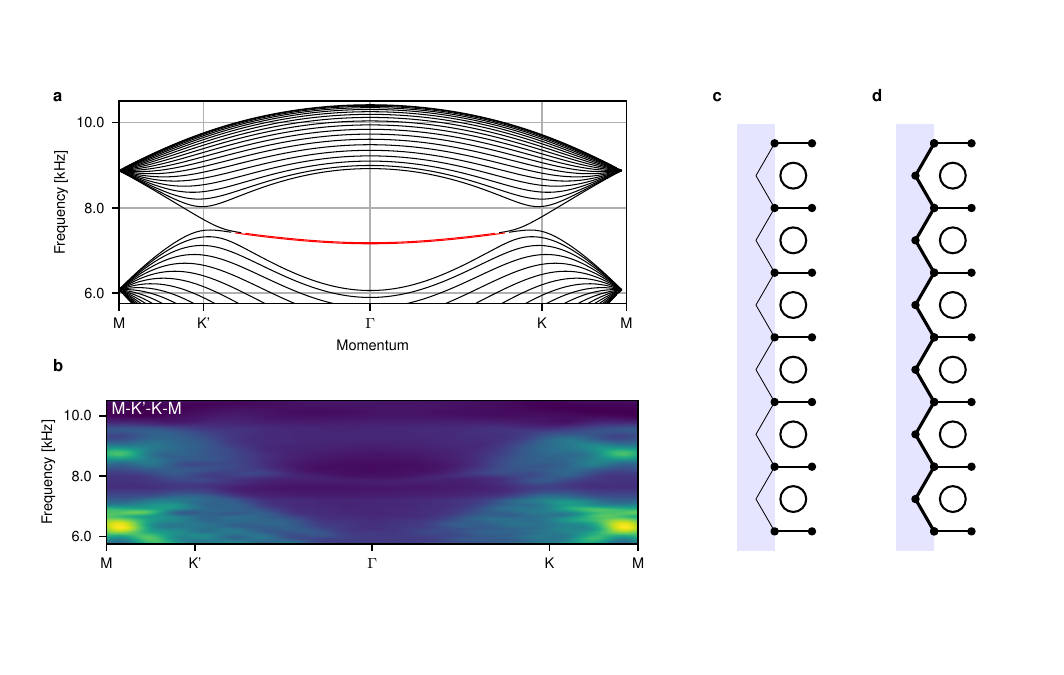}
\caption{{\bf Open boundaries on $\pm\hat{\bf x}$ surfaces.} {\bf a,} Band structure of the tight binding model as in Fig.~\ref{fig:suppTBField}. The only difference is that the surfaces $\pm\hat{\bf x}$ have bearded edge terminations rather than zigzag. the edge modes in red are located at $k_y\in[-2\pi/3\,,2\pi/3]$. {\bf b,} Measured spectrum along $k_y$ for the metamaterial with open boundaries at the $\pm\hat{\bf x}$ surfaces. Compared to Fig.2d of the main text, extra modes in the gap for $k_y\in[-2\pi/3\,,2\pi/3]$ appear. These modes are compatible with surface modes for bearded edges. {\bf c,} Schematic representation of the $-\hat{\bf x}$ surface with open boundaries. The blue area represent the open space. The acoustic pressure field inside the sample does not see the last point of the sonic crystal that belong to open space. This results in an effective bearded edge as supported by {\bf b}. {\bf d,} Schematic representation of a zigzag termination. The zigzag boundary is shown with a thicker line. The separation between sonic crystal and open space is now clear and the acoustic pressure field inside the sample experiences a zigzag edge. }
\label{fig:suppBoundary}
\end{figure}
\fi%

On the other hand, leaving the sample open would alter the termination on those surfaces. In fact, Fig.~\ref{fig:suppBoundary}b shows the measured spectrum for the open sample on $\pm\hat{\bf x}$. Modes that are not present in the closed sample measurements appear in the range $k_y\in [-2\pi/3\,,2\pi/3]$ and in the frequency range of the chiral channels. These are compatible with modes of a bearded edge surface termination. The presence of these modes for the open sample can be intuitively understood as follows. In the open case, the most extremal lattice site of the hexagonal unit cell at the boundary is decoupled from the rest of the crystal. In fact, the acoustic pressure field feels the presence of the last pillar, but at the most extremal lattice point it is effectively radiating in free space. This results in an effective bearded edge termination for the sound field. A schematic representation of this situation is portrayed in Fig.~\ref{fig:suppBoundary}c--d.

\section{ Real space data}
\label{sec:flat}
\noindent
The evidences for the axial field induced chiral channels rely on momentum space observations. For completeness, we provide in Fig.~\ref{fig:suppRealSpace} some examples of real space measurements of the acoustic pressure field. 

For each panel we show the logarithm of the absolute value of the pressure field measured along two coordinates. The other coordinate not shown is summed over. Each column displays a different 2D plane. All the data refers to excitations at the channel frequency, $7.7 {\rm kHz}$. From these figures the excitation position for each dataset can be readily determined.

There is an asymmetry between the propagation along  $\hat{\bf z}$ with respect to $\hat{\bf x}$ and $\hat{\bf y}$. This is due to the ratio $t_n/t_c\approx 14.1$ discussed above. The axis coordinates are in units of measured sites. The actual physical length of the sample along the $x$ direction is $39.7 { \rm cm}$, and $45 {\rm cm}$ along $y$. 

It is hard to argue for an imbalance between the propagation along $y$ and $x$. This can be understood by the width of the flat region of the zeroth Landau level in the direction perpendicular to the field. It is mainly affected by the number of unit cells in the direction along which the gauge potential varies. In our case, the window in momentum and frequency where we expect a flat Landau level is narrower than what our quality factor allows us to resolve. A similar procedure to the phase fitting of the chiral channels cannot be applied to the flat Landau level since at fixed frequency there are multiple $k_z$ values. 

These are the main reasons why it is not possible to strongly argue for an imbalance in the $x$ and $y$ direction propagation from real space data. In particular, tight binding simulations, shown in Fig~\ref{fig:suppTBFlat}, for parameter values extrapolated from the metamaterial's ones allows to study how the size of the flat Landau level changes as a function of $L_x$. At fixed value of the axial gauge field $\tilde B_5=0.08$, for $L_x=20$ we have a flat Landau level for $\Delta k_z \approx 0.4$ and for $L_x=40$ we have $\Delta k_z \approx 1.1$.
Note that the resolution of the Fourier transformed data in our experiment is $\delta k_z = 0.5$. Hence, the tight binding model predicts a flat regions below the measurement resolution. This also prevents us from studying the position-momentum locking of the eigenstates of the zeroth flat Landau level. 

\ifpreprint%
\else%
\begin{figure}[tb]
\includegraphics{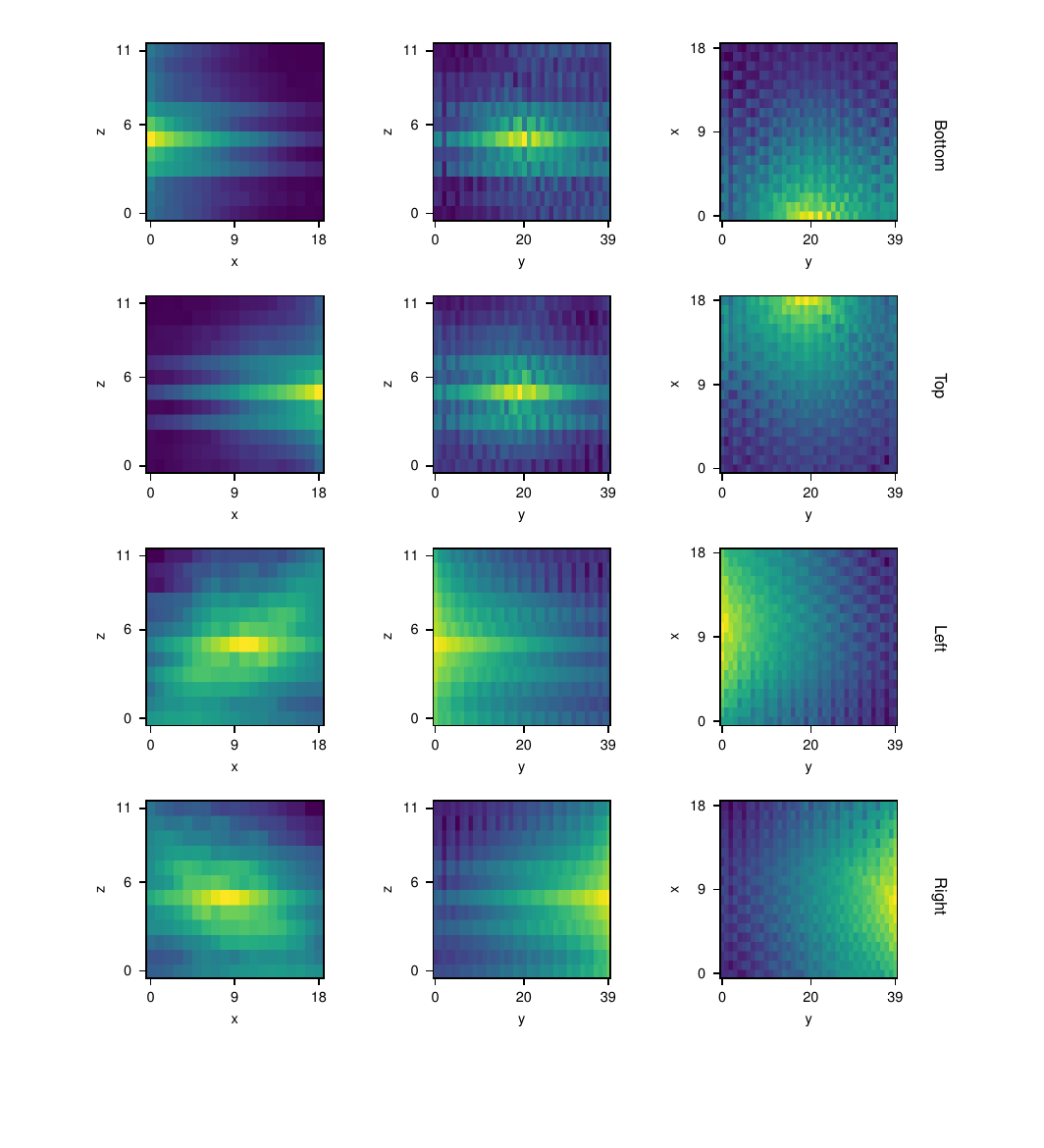}
\caption{{\bf Real space pressure field data.} 2D images of the logarithm of the absolute value of the real space measured acoustic pressure field. The coordinate not shown is summed over and the axes are in measurement point units. The mode at $7.7 {\rm kHz}$ is displayed. In the first column the $xz$ plane is shown. In the second, the $yz$ plane and in the last column the $xy$ one. Each row displays excitation from a different surface as indicated by the labels on the right. From the picture the location of the speaker can be determined. The clear imbalance between propagation along $z$ and the other directions is due to the asymmetry between the in-plane and out-of-plane couplings. Due to dissipations in our system, a clear imbalance in the propagation along $x$ and $y$ cannot be observed in the real space data.}
\label{fig:suppRealSpace}
\end{figure}
\fi%

\ifpreprint%
\else%
\begin{figure}[tb]
\includegraphics{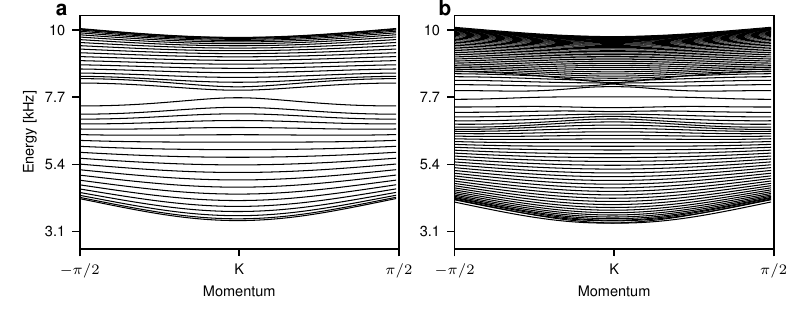}
\caption{{\bf Flatness of zeroth Landau level.} The band structure for the tight binding model in an axial gauge field $\tilde B_5=0.08$ along the direction perpendicular to the field. In both a zeroth flat Landau level can be seen at $K$. {\bf a,} Sample of 20 unit cells along $x$. The flat level spans a momentum interval $\Delta k_z \approx 0.4$. {\bf b,} Sample of 40 unit cells along $x$ and same field as in {\bf a}. The flat level spans a momentum interval $\Delta k_z \approx 1.1$. }
\label{fig:suppTBFlat}
\end{figure}
\fi%

\section{ Mode-localization in the 2D Brillouin zone}
\noindent
In the main text, the localization of the modes inside the 2D Brillouin zone has been provided as an evidence of the zeroth Landau levels chirality. However, measurements for only few frequencies have been provided there. Here, more data are displayed in Fig.~\ref{fig:suppBZ}.

\ifpreprint%
\else%
\begin{figure}[tb]
\includegraphics{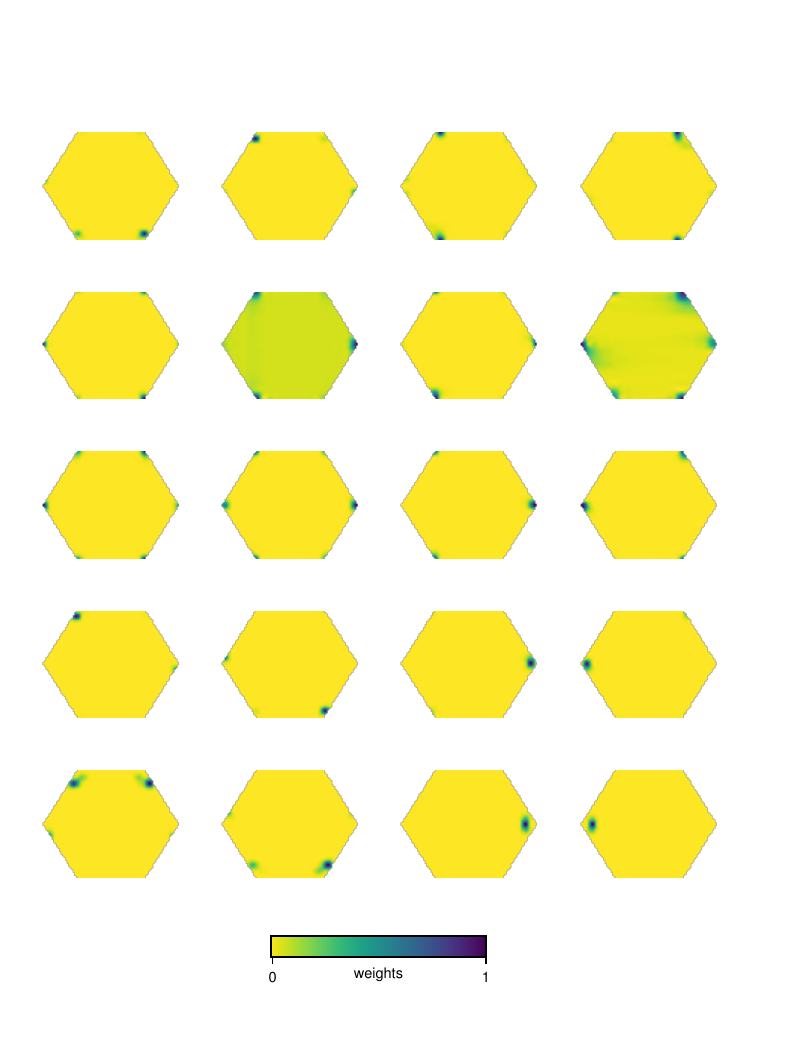}
\caption{{\bf Modes localization in 2D Brillouin zone.}  
Fourier transform to the first Brillouin zone at $k_z=0$ of the acoustic response at different frequencies for different speaker positions (top \& bottom, right, or left, see Main Text). Small frequency steps around the chiral channel frequency $7.7 {\rm kHz}$ are shown. }
\label{fig:suppBZ}
\end{figure}
\fi%

A video evolutionBZ.mp4 showing the localization of the modes in the 2D BZ at all the measured frequencies is available as Supplementary material.

\section{Fermi arcs localization}
\noindent
Fermi arcs are the boundary manifestation of the non-trivial bulk topology of Weyl systems. They are open equi-energy surfaces on the 2D surface of the 3D Weyl system. The presence of open equi-frequency surfaces is a peculiar feature allowed by the presence of the 3D bulk. At the momentum projection of the Weyl points over the 2D surface, bulk and boundary modes mix together. A striking feature of Fermi arcs is hence a momentum dependent decay length of the surface states. For momenta away from the Weyl point projection, the states are well localized on the surface. On the other hand, at the projection momentum, the decay length grows and the distinction between bulk and surface modes is ill-defined. We measure the decay length of the Fermi arcs in our sample and confirme this. 

In particular, we focus the attention on the $-\hat{\bf x}$ (bottom) surface. We take the real space pressure field data and perform discrete Fourier transform along $\hat{\bf z}$. The decay of the surface modes along $\hat{\bf x}$ as they propagate in the bulk, is studied summing over all the $\hat{\bf y}$ components. The decay is then fitted to an exponential in order to obtain the decay length and its uncertainty. Fig.~\ref{fig:suppArcs} shows the measured decay of the Fermi arcs at frequency $7.7 {\rm kHz}$ for momentum $k_z$ equal to 0 ($\pi/2$) in violet (green).

\ifpreprint%
\else%
\begin{figure}[tb]
\includegraphics{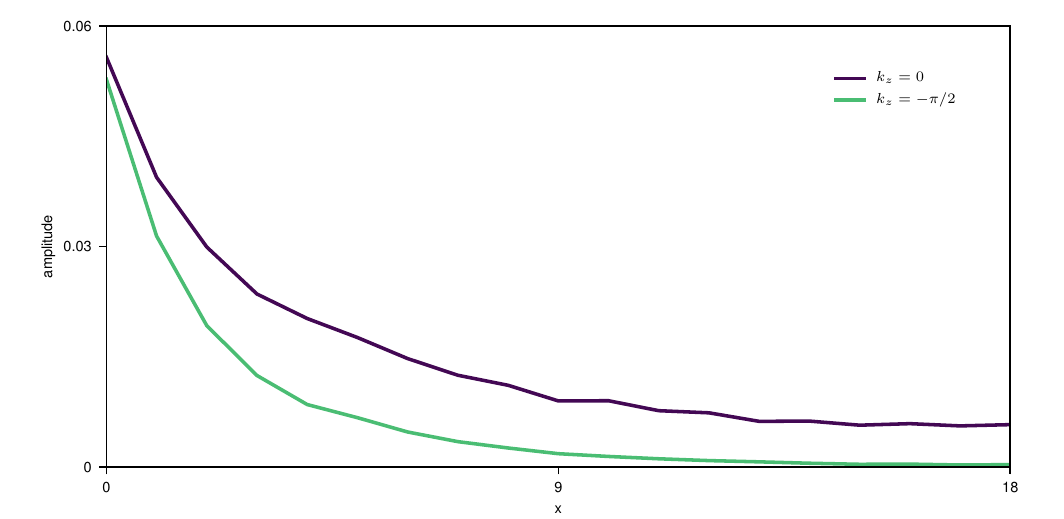}
\caption{{\bf Surface Fermi arcs localization.} Decay of the modes on the $-\hat{\bf x}$ surface at $7.7\, {\rm kHz}$ at momenta $k_z=0$ (violet) and $k_z=\pi/2$ (green). At $k_z=0$, the WPs project on the surface 2D BZ and bulk and boundary modes mix. At $k_z=\pi/2$, the Fermi arcs are well localized on the surface. At momenta where bulk and boundary modes mix the measured pressure field is significantly different from zero through the whole sample.}
\label{fig:suppArcs}
\end{figure}
\fi%

From the fit, we obtained a decay length of $3.22$ unit cells ($6.3 {\rm cm}$) at $k_z=0$ and $2.05$ ($4.0 {\rm cm}$) at $k_z=\pi/2$. The uncertainty obtained from the fit is in both cases $0.02$ unit cells ($0.4 {\rm cm}$). The measured decay is similar on the other surface and for other momenta and confirms the longer decay length for surface modes directly connected to bulk modes at the projection of WPs on the 2D surface BZ. Note also how for $k_z=0$ the response is significantly different from zero across the whole sample. This proves the mixed bulk-boundary nature of this mode.

\section{Single unit cells parameters}
The unit cell of the phononic crystal is shown in Fig.1c of the main text. To implement an axial gauge field, we design a gradient along the $x$-direction by introducing holes above one of the two sub-lattices with varying radius $n$ and fixed depth $m$. This gradient shifts the WPs along $k_z$ as a function of $x$, implementing the main ingredient for the realization of an axial gauge field. However, such a space dependent perturbation could introduce a tilt in the Weyl cones, a shift in frequency and other undesired effects. The geometric parameters of each unit cell ($h$, $R$, $n$) are adjusted via full wave simulations to reduce these undesired changes in the Weyl dispersion. The final crystal is then obtained stacking the optimised unit cells along the $x$-direction and hence giving rise to the sought after axial gauge filed. The velocities obtain for each unit cell via full wave simulations are presented in Table~\ref{tab:geoVelo}.

Note here: The group velocity in $z$-direction fluctuate considerably from unit-cell to unit-cell simulated here. However, the measurements of $v_z$ presented above yields a relatively small error-bar. We conclude that via the assembly of the single-unit cells to a finite sample, we obtain a well defined average $v_z$.

\begin{table}[h!]
\centering
\begin{tabular}{l | c  c  c | c c  }
   & $h$  &  $R$ &  $n$  & $\tilde v_n$  & $\tilde v_z$  \\
   &($\rm mm$) & ($\rm mm$) &($\rm mm$) & ($\rm m/s$) & ($\rm m/s$)\\ 
	\hline
	 UC 1  & 8.5 & 6.5 & 2.6 & 26.8 & 7.3 \\
         UC 2  & 8.0 & 7.3 & 1.9 & 29.2 & 4.8  \\
         UC 3  & 8.3 & 6.9 & 2.1 & 28.2 & 6.4  \\
         UC 4  & 8.5 & 6.8 & 2.0 & 27.8 & 7.1  \\
         UC 5  & 8.7 & 6.8 & 1.9 & 27.5 & 7.6  \\
         UC 6  & 8.3 & 7.4 & 1.3 & 29.3 & 5.3  \\
         UC 7  & 8.3 & 7.5 & 1.1 & 29.6 & 4.8  \\
         UC 8  & 8.9 & 6.9 & 1.2 & 27.7 & 7.8  \\
         UC 9  & 9.0 & 7.2 & 0.4 & 29.3 & 13.3 \\
         UC 10 & 9.3 & 6.8 & 0.0 & 26.7 & 8.9  \\
         UC 11 & 8.9 & 7.4 & 0.4 & 29.8 & 11.3\\
         UC 12 & 8.7 & 7.2 & 1.0 & 28.7 & 6.4 \\
         UC 13 & 8.2 & 7.6 & 1.1 & 29.8 & 4.5 \\
         UC 14 & 8.4 & 7.3 & 1.4 & 29.1 & 5.7 \\
         UC 15 & 8.7 & 6.8 & 1.9 & 27.5 & 7.6 \\
         UC 16 & 8.5 & 6.8 & 2.0 & 27.7 & 7.0 \\
         UC 17 & 8.3 & 6.9 & 2.1 & 28.2 & 6.6 \\
         UC 18 & 8.0 & 7.3 & 1.9 & 29.2 & 4.9 \\
         UC 19 & 8.5 & 6.5 & 2.6 & 26.9 & 7.3 \\
         UC 20 & 8.7 & 6.1 & 3.0 & 27.0 & 8.0 \\
	\hline
	\end{tabular}
  \caption{\label{tab:geoVelo}Values of the geometrical parameters and velocities along the linear dispersions for each unit cell (UC) of the acoustic crystal. The WP frequencies are fixed to 7.7 kHz by design.}
\end{table}	

\section{Exceptional rings and axial field}
One may wonder whether the presence of strong dissipation alters the physics that describes our system. In particular, the appearance of exceptional rings may have a detrimental effect on the chiral channels induced by an axial field. Our experimental results suggest that the chiral channels survive our specific dissipation. Moreover,  we argue here on theoretical grounds why this is the case.

A simple example of non-Hermitian system with exceptional rings is described by the Hamiltonian
\begin{equation}
  \label{dissH}
  H({\bf k})=\sum_{j=1}^3k_j\sigma^j+i \gamma\sigma^z\,,
\end{equation}
where $\sigma^j$ are the Pauli matrices and $\gamma$ is a dissipative term that can be associated to a sublattice dependent gain/loss process. The spectrum of the above Hamiltonian is given by
\begin{equation}
  \epsilon({\bf k})=\sqrt{{\bf k}^2-\gamma^2+2ik_z\gamma}\,.
\end{equation}
When ${\bf k}^2=\gamma^2$ and $k_z=0$ both the real and imaginary parts of the eigenvalues are zero: The Weyl point turned into an exceptional ring after the introduction of a non-Hermitian term. Such exceptional ring have recently attracted much interest and have been studied in detail, see, e.g. Refs.~[\onlinecite{Xu17z,Cerjan18z}].

The main source of dissipation in our system is uniform and enters the Hamiltonian via the identity matrix. This decreases the quality factor and hence blurs all the spectral features. However, as it does not couple to one of the three Pauli matrices, this term does not induce exceptional rings.

A priori, we cannot exclude the presence of other dissipative terms. In particular, it is reasonable to assume a dissipative $\sigma^z$ term. Acoustic losses occur as a consequence of surface roughness. The surface to volume ratio is particularly relevant for the small sublattice notches and ventilator holes. Both these components are captured by a $\sigma^z$ term in the tight binding Hamiltonian which describes our system around the Weyl points. The axial gauge potential couples via that Pauli matrix too and it introduces a gradient along the $x$-direction. This gradient induces changes to both the real-part of the eigenvalues as well as to the dissipation. It is thus reasonable to study a dissipative term of the form $\propto i\gamma x \sigma^z $, with $\gamma \in \mathbb{R}$.

We consider a Hamiltonian similar to Eq.~\eqref{dissH}
\begin{equation}
  H({\bf k})=\sum_{j=1}^3\tilde{k}_j\sigma^j+i \gamma x\sigma^z\,,
\end{equation}
where the dissipative term is now spatially dependent and $\tilde{{\bf k}}=(k_x\,,k_y ,k_z+Bx)$. The gauge potential enters the Hamiltonian via minimal coupling. We apply the usual procedure to find relativistic Landau levels and consider the spectrum of the Hamiltonian squared:
\begin{equation}
\epsilon({\bf k})^2=[k_z+(B+i\gamma)x]^2+k_x^2+k_y^2 \pm (B+i\gamma)\,.
\end{equation}
This spectrum is similar to the one obtained in the absence of a dissipative term. The only difference it that the field $B$ is replaced by the complex field $B^*=B+i\gamma$. In particular, the functional form of the eigenvectors is unchanged and the chiral LL follows from the procedure presented in Ref.~\onlinecite{Nielsen83}. The spectrum of the Hamiltonian in the presence of an axial gauge field and a spatially dependent non-Hermitian term is:
\begin{equation}
  \epsilon({\bf k})=\pm\sqrt{k_y^2+2n(B+i\gamma)}\,.
  \end{equation}
The chiral LL level ($n=0$) does not seem to be affected by the extra dissipative term. On the other hand, the complex term does not give rise to exceptional ring since the real and imaginary part are never simultaneously zero.

	\end{bibunit}


\begin{thebibliography}{10}
\expandafter\ifx\csname url\endcsname\relax
  \def\url#1{\texttt{#1}}\fi
\expandafter\ifx\csname urlprefix\endcsname\relax\def\urlprefix{URL }\fi
\providecommand{\bibinfo}[2]{#2}
\providecommand{\eprint}[2][]{\url{#2}}

\bibitem{Roy18}
\bibinfo{author}{Roy, S.}, \bibinfo{author}{Kolodrubetz, M.},
  \bibinfo{author}{Goldman, N.} \& \bibinfo{author}{Grushin, A.~G.}
\newblock \bibinfo{title}{Tunable axial gauge fields in engineered {Weyl}
  semimetals: semiclassical analysis and optical lattice implementations}.
\newblock \emph{\bibinfo{journal}{2D Materials}} \textbf{\bibinfo{volume}{5}},
  \bibinfo{pages}{024001} (\bibinfo{year}{2018}).
\newblock \urlprefix\url{https://doi.org/10.1088/2053-1583/aaa577}.

\bibitem{Lu15}
\bibinfo{author}{Lu, L.} \emph{et~al.}
\newblock \bibinfo{title}{Experimental observation of {Weyl} points}.
\newblock \emph{\bibinfo{journal}{Science}} \textbf{\bibinfo{volume}{349}},
  \bibinfo{pages}{6248} (\bibinfo{year}{2015}).
\newblock \urlprefix\url{https://dx.doi.org/10.1126/science.aaa9273}.

\bibitem{Li17}
\bibinfo{author}{Li, F.}, \bibinfo{author}{Huang, X.}, \bibinfo{author}{Lu,
  J.}, \bibinfo{author}{Ma, J.} \& \bibinfo{author}{Liu, Z.}
\newblock \bibinfo{title}{Weyl points and fermi arcs in a chiral phononic
  crystal}.
\newblock \emph{\bibinfo{journal}{Nature Phys.}}  (\bibinfo{year}{2017}).
\newblock \urlprefix\url{https://dx.doi.org/10.1038/nphys4275}.

\bibitem{Mourik12}
\bibinfo{author}{Mourik, V.} \emph{et~al.}
\newblock \bibinfo{title}{Signatures of {Majorana} fermions in hybrid
  superconductor-semiconductor nanowire devices}.
\newblock \emph{\bibinfo{journal}{Science}} \textbf{\bibinfo{volume}{336}},
  \bibinfo{pages}{1003} (\bibinfo{year}{2012}).
\newblock \urlprefix\url{http://dx.doi.org/10.1126/science.1222360}.

\bibitem{Borisenko14}
\bibinfo{author}{Borisenko, S.} \emph{et~al.}
\newblock \bibinfo{title}{Experimental realization of a three-dimensional
  {Dirac} semimetal}.
\newblock \emph{\bibinfo{journal}{Phys. Rev. Lett.}}
  \textbf{\bibinfo{volume}{113}}, \bibinfo{pages}{027603}
  (\bibinfo{year}{2014}).
\newblock \urlprefix\url{https://doi.org/10.1103/PhysRevLett.113.027603}.

\bibitem{Liu14a}
\bibinfo{author}{Liu, Z.~K.} \emph{et~al.}
\newblock \bibinfo{title}{Discovery of a three-dimensional topological dirac
  semimetal, {Na$_3$Bi}}.
\newblock \emph{\bibinfo{journal}{Science}} \textbf{\bibinfo{volume}{343}},
  \bibinfo{pages}{864} (\bibinfo{year}{2014}).
\newblock \urlprefix\url{https://dx.doi.org/10.1126/science.1245085}.

\bibitem{Weng15}
\bibinfo{author}{Weng, H.}, \bibinfo{author}{Fang, C.}, \bibinfo{author}{Fang,
  Z.}, \bibinfo{author}{Bernevig, B.~A.} \& \bibinfo{author}{Dai, X.}
\newblock \bibinfo{title}{Weyl semimetal phase in noncentrosymmetric
  transition-metal monophosphides}.
\newblock \emph{\bibinfo{journal}{Phys. Rev. X}} \textbf{\bibinfo{volume}{5}},
  \bibinfo{pages}{011029} (\bibinfo{year}{2015}).
\newblock \urlprefix\url{https://doi.org/10.1103/PhysRevX.5.011029}.

\bibitem{Xu15}
\bibinfo{author}{Xu, S.-Y.} \emph{et~al.}
\newblock \bibinfo{title}{Discovery of a {Weyl} fermion semimetal and
  topological {Fermi} arcs}.
\newblock \emph{\bibinfo{journal}{Science}} \textbf{\bibinfo{volume}{349}},
  \bibinfo{pages}{6248} (\bibinfo{year}{2015}).
\newblock \urlprefix\url{https://dx.doi.org/10.1126/science.aaa9297}.

\bibitem{Lv15}
\bibinfo{author}{Lv, B.} \emph{et~al.}
\newblock \bibinfo{title}{Experimental discovery of {Weyl} semimetal {TaAs}}.
\newblock \emph{\bibinfo{journal}{Phys. Rev. X}} \textbf{\bibinfo{volume}{5}},
  \bibinfo{pages}{031013} (\bibinfo{year}{2015}).
\newblock \urlprefix\url{https://doi.org/10.1103/PhysRevX.5.031013}.

\bibitem{Weyl29}
\bibinfo{author}{Weyl, H.}
\newblock \bibinfo{title}{{Elektron und Gravitation. I}}.
\newblock \emph{\bibinfo{journal}{Z. Phys.}} \textbf{\bibinfo{volume}{56}},
  \bibinfo{pages}{330} (\bibinfo{year}{1929}).
\newblock \urlprefix\url{https://doi.org/10.1007/BF01339504}.

\bibitem{Adler69}
\bibinfo{author}{Adler, S.~L.}
\newblock \bibinfo{title}{Axial-vector vertex in spinor electrodynamics}.
\newblock \emph{\bibinfo{journal}{Phys. Rev.}} \textbf{\bibinfo{volume}{177}},
  \bibinfo{pages}{2426} (\bibinfo{year}{1969}).
\newblock \urlprefix\url{https://doi.org/10.1103/PhysRev.177.2426}.

\bibitem{Bell69}
\bibinfo{author}{Bell, J.~S.} \& \bibinfo{author}{Jackiw, R.}
\newblock \bibinfo{title}{A {PCAC} puzzle: $\pi_0\rightarrow \gamma\gamma$ in
  the $\sigma$-model}.
\newblock \emph{\bibinfo{journal}{Nuovo Cimento}}
  \textbf{\bibinfo{volume}{60}}, \bibinfo{pages}{47} (\bibinfo{year}{1969}).
\newblock \urlprefix\url{https://doi.org/10.1007/BF02823296}.

\bibitem{Bertlmann00}
\bibinfo{author}{Bertlmann, R.~A.}
\newblock \emph{\bibinfo{title}{Anomalies in Quantum Field Theory}}
  (\bibinfo{publisher}{Oxford University Press}, \bibinfo{address}{Oxford},
  \bibinfo{year}{2000}).

\bibitem{Landsteiner16}
\bibinfo{author}{Landsteiner, K.}
\newblock \bibinfo{title}{Notes on anomaly induced transport}.
\newblock \emph{\bibinfo{journal}{Acta Phys. Pol. B}}
  \textbf{\bibinfo{volume}{47}}, \bibinfo{pages}{2617} (\bibinfo{year}{2016}).
\newblock \urlprefix\url{https://dx.doi.org/10.5506/APhysPolB.47.2617}.

\bibitem{Gooth17}
\bibinfo{author}{Gooth, J.} \emph{et~al.}
\newblock \bibinfo{title}{Experimental signatures of the mixed
  axial--gravitational anomaly in the {Weyl} semimetal {NbP}}.
\newblock \emph{\bibinfo{journal}{Nature}} \textbf{\bibinfo{volume}{547}},
  \bibinfo{pages}{324} (\bibinfo{year}{2017}).
\newblock \urlprefix\url{https://dx.doi.org/10.1038/nature23005}.

\bibitem{Pikulin16}
\bibinfo{author}{Pikulin, D.~I.}, \bibinfo{author}{Chen, A.} \&
  \bibinfo{author}{Franz, M.}
\newblock \bibinfo{title}{Chiral anomaly from strain-induced gauge fields in
  {Dirac} and {Weyl} semimetals}.
\newblock \emph{\bibinfo{journal}{Phys. Rev. X}} \textbf{\bibinfo{volume}{6}},
  \bibinfo{pages}{041021} (\bibinfo{year}{2016}).
\newblock \urlprefix\url{https://dx.doi.org/10.1103/PhysRevX.6.041021}.

\bibitem{Potter14}
\bibinfo{author}{Potter, A.~C.}, \bibinfo{author}{Kimchi, I.} \&
  \bibinfo{author}{Vishwanath, A.}
\newblock \bibinfo{title}{Quantum oscillations from surface {Fermi} arcs in
  {Weyl} and {Dirac} semimetals}.
\newblock \emph{\bibinfo{journal}{Nature Comm.}} \textbf{\bibinfo{volume}{5}},
  \bibinfo{pages}{5161} (\bibinfo{year}{2014}).
\newblock \urlprefix\url{https://dx.doi.org/10.1038/ncomms6161}.

\bibitem{Huber16}
\bibinfo{author}{Huber, S.~D.}
\newblock \bibinfo{title}{Topological mechanics}.
\newblock \emph{\bibinfo{journal}{Nature Phys.}} \textbf{\bibinfo{volume}{12}},
  \bibinfo{pages}{621} (\bibinfo{year}{2016}).
\newblock \urlprefix\url{http://dx.doi.org/10.1038/nphys3801}.

\bibitem{Nielsen83}
\bibinfo{author}{Nielsen, H.} \& \bibinfo{author}{Ninomia, M.}
\newblock \bibinfo{title}{The {Adler-Bell-Jackiw} anomaly and {Weyl} fermions
  in a crystal}.
\newblock \emph{\bibinfo{journal}{Phys. Lett. B}}
  \textbf{\bibinfo{volume}{130}}, \bibinfo{pages}{389} (\bibinfo{year}{1983}).
\newblock \urlprefix\url{https://doi.org/10.1016/0370-2693(83)91529-0}.

\bibitem{Bernevig15}
\bibinfo{author}{Bernevig, B.~A.}
\newblock \bibinfo{title}{It's been a {Weyl} coming}.
\newblock \emph{\bibinfo{journal}{Nature Phys.}} \textbf{\bibinfo{volume}{11}},
  \bibinfo{pages}{698} (\bibinfo{year}{2015}).
\newblock \urlprefix\url{http://dx.doi.org/10.1038/nphys3454}.

\bibitem{Yang15a}
\bibinfo{author}{Yang, L.} \emph{et~al.}
\newblock \bibinfo{title}{{Weyl} semimetal phase in the non-centrosymmetric
  compound {TaAs}}.
\newblock \emph{\bibinfo{journal}{Nature Phys.}} \textbf{\bibinfo{volume}{11}},
  \bibinfo{pages}{728} (\bibinfo{year}{2015}).
\newblock \urlprefix\url{https://dx.doi.org/10.1038/nphys3425}.

\bibitem{Moll16}
\bibinfo{author}{Moll, P. J.~W.} \emph{et~al.}
\newblock \bibinfo{title}{Transport evidence for {Fermi-arc-mediated} chirality
  transfer in the {Dirac} semimetal {Cd$_3$As$_2$}}.
\newblock \emph{\bibinfo{journal}{Nature}} \textbf{\bibinfo{volume}{535}},
  \bibinfo{pages}{266} (\bibinfo{year}{2016}).
\newblock \urlprefix\url{https://dx.doi.org/10.1038/nature18276}.

\bibitem{Klitzing80}
\bibinfo{author}{v.~Klitzing, K.}, \bibinfo{author}{Dorda, G.} \&
  \bibinfo{author}{Pepper, M.}
\newblock \bibinfo{title}{New method for high-accuracy determination of the
  fine-structure constant based on quantized {Hall} resistance}.
\newblock \emph{\bibinfo{journal}{Phys. Rev. Lett.}}
  \textbf{\bibinfo{volume}{45}}, \bibinfo{pages}{494} (\bibinfo{year}{1980}).
\newblock \urlprefix\url{http://link.aps.org/doi/10.1103/PhysRevLett.45.494}.

\bibitem{Xiao15b}
\bibinfo{author}{Xiao, M.}, \bibinfo{author}{Chen, W.-J.}, \bibinfo{author}{He,
  W.-Y.} \& \bibinfo{author}{Chan, C.~T.}
\newblock \bibinfo{title}{Synthetic gauge flux and {Weyl} points in acoustic
  systems}.
\newblock \emph{\bibinfo{journal}{Nature Phys.}} \textbf{\bibinfo{volume}{11}},
  \bibinfo{pages}{920} (\bibinfo{year}{2015}).
\newblock \urlprefix\url{http://dx.doi.org/10.1038/nphys3458}.

\bibitem{He18}
\bibinfo{author}{He, H.} \emph{et~al.}
\newblock \bibinfo{title}{Topological negative refraction of surface acoustic
  waves in a weyl phononic crystal}.
\newblock \emph{\bibinfo{journal}{Nature}} \textbf{\bibinfo{volume}{560}},
  \bibinfo{pages}{61} (\bibinfo{year}{2018}).
\newblock \urlprefix\url{https://doi.org/10.1038/s41586-018-0367-9}.

\bibitem{Vozmediano10}
\bibinfo{author}{Vozmediano, M. A.~H.}, \bibinfo{author}{Katsnelson, M.~I.} \&
  \bibinfo{author}{Guinea, F.}
\newblock \bibinfo{title}{Gauge fields in graphene}.
\newblock \emph{\bibinfo{journal}{Phys. Rep.}} \textbf{\bibinfo{volume}{496}},
  \bibinfo{pages}{109} (\bibinfo{year}{2010}).
\newblock \urlprefix\url{https://dx.doi.org/10.1016/j.physrep.2010.07.003}.

\bibitem{Liu13}
\bibinfo{author}{Liu, C.-X.}, \bibinfo{author}{Ye, P.} \& \bibinfo{author}{Qi,
  X.-L.}
\newblock \bibinfo{title}{Chiral gauge field and axial anomaly in a {Weyl}
  semimetal}.
\newblock \emph{\bibinfo{journal}{Phys. Rev. B}} \textbf{\bibinfo{volume}{87}},
  \bibinfo{pages}{235306} (\bibinfo{year}{2013}).
\newblock \urlprefix\url{https://dx.doi.org/10.1103/PhysRevB.87.235306}.

\bibitem{Cortjio15}
\bibinfo{author}{Cortjio, A.}, \bibinfo{author}{Ferreir{\'o}s, Y.},
  \bibinfo{author}{Landsteiner, K.} \& \bibinfo{author}{Vozmediano, M. A.~H.}
\newblock \bibinfo{title}{Elastic gauge fields in {Weyl} semimetals}.
\newblock \emph{\bibinfo{journal}{Phys. Rev. Lett.}}
  \textbf{\bibinfo{volume}{115}}, \bibinfo{pages}{177202}
  (\bibinfo{year}{2015}).
\newblock \urlprefix\url{https://doi.org/10.1103/PhysRevLett.115.177202}.

\bibitem{Grushin16}
\bibinfo{author}{Grushin, A.~G.}, \bibinfo{author}{Venderbros, J. W.~F.},
  \bibinfo{author}{Vishwanath, A.} \& \bibinfo{author}{Ilan, R.}
\newblock \bibinfo{title}{Inhomogeneous {Weyl} and {Dirac} semimetals:
  Transport in axial magnetic fields and {Fermi} arc surface states from
  pseudo-{Landau} levels}.
\newblock \emph{\bibinfo{journal}{Phys. Rev. X}} \textbf{\bibinfo{volume}{6}}
  (\bibinfo{year}{2016}).
\newblock \urlprefix\url{https://doi.org/10.1103/PhysRevX.6.041046}.

\bibitem{Sumiyoshi16}
\bibinfo{author}{Sumiyoshi, H.} \& \bibinfo{author}{Fujimoto, S.}
\newblock \bibinfo{title}{Torsional chiral magnetic effect in a {Weyl}
  semimetal with a topological defect}.
\newblock \emph{\bibinfo{journal}{Phys. Rev. Lett.}}
  \textbf{\bibinfo{volume}{116}}, \bibinfo{pages}{166601}
  (\bibinfo{year}{2016}).
\newblock \urlprefix\url{https://doi.org/10.1103/PhysRevLett.116.166601}.

\bibitem{Abbaszadeh17}
\bibinfo{author}{Abbaszadeh, H.}, \bibinfo{author}{Souslov, A.},
  \bibinfo{author}{Paulose, J.}, \bibinfo{author}{Schomerus, H.} \&
  \bibinfo{author}{Vitelli, V.}
\newblock \bibinfo{title}{Sonic landau-level lasing and synthetic gauge fields
  in mechanical metamaterials}.
\newblock \emph{\bibinfo{journal}{Phys. Rev. Lett.}}
  \textbf{\bibinfo{volume}{119}}, \bibinfo{pages}{195502}
  (\bibinfo{year}{2017}).
\newblock \urlprefix\url{https://doi.org/10.1103/PhysRevLett.119.195502}.

\bibitem{Levy10}
\bibinfo{author}{Levy, N.} \emph{et~al.}
\newblock \bibinfo{title}{Strain-induced pseudo--magnetic fields greater than
  300 {Tesla} in graphene nanobubbles}.
\newblock \emph{\bibinfo{journal}{Science}} \textbf{\bibinfo{volume}{329}},
  \bibinfo{pages}{544} (\bibinfo{year}{2010}).
\newblock \urlprefix\url{https://dx.doi.org/10.1126/science.1191700}.

\bibitem{Ni18}
\bibinfo{author}{Ni, X.}, \bibinfo{author}{Weiner, M.},
  \bibinfo{author}{Al{\`u}, A.} \& \bibinfo{author}{Khanikaev, A.~B.}
\newblock \bibinfo{title}{Observation of bulk polarization transitions and
  higher-order embedded topological eigenstates for sound}.
\newblock \emph{\bibinfo{journal}{arXiv:1807.00896}}  (\bibinfo{year}{2018}).
\newblock \urlprefix\url{https://arxiv.org/abs/1807.00896}.

\bibitem{Xu17z}
\bibinfo{author}{Xu, Y.}, \bibinfo{author}{Wang, S.-T.} \&
  \bibinfo{author}{Duan, L.-M.}
\newblock \bibinfo{title}{Weyl exceptional rings in a three-dimensional
  dissipative cold atomic gas}.
\newblock \emph{\bibinfo{journal}{Phys. Rev. Lett.}}
  \textbf{\bibinfo{volume}{118}}, \bibinfo{pages}{045701}
  (\bibinfo{year}{2017}).
\newblock
  \urlprefix\url{https://link.aps.org/doi/10.1103/PhysRevLett.118.045701}.

\bibitem{Yang17}
\bibinfo{author}{Yang, Z.}, \bibinfo{author}{Gao, Z.}, \bibinfo{author}{Yang,
  Y.} \& \bibinfo{author}{Zhang, B.}
\newblock \bibinfo{title}{Strain-induced gauge field and landau levels in
  acoustic structures}.
\newblock \emph{\bibinfo{journal}{Phys. Rev. Lett.}}
  \textbf{\bibinfo{volume}{118}}, \bibinfo{pages}{194301}
  (\bibinfo{year}{2017}).
\newblock
  \urlprefix\url{https://journals.aps.org/prl/abstract/10.1103/PhysRevLett.118.194301}.

\bibitem{Wen18}
\bibinfo{author}{Wen, X.} \emph{et~al.}
\newblock \bibinfo{title}{Observation of acoustic landau quantization and
  quantum-hall-like edge states}.
\newblock \emph{\bibinfo{journal}{arXiv:1807.08454}}  (\bibinfo{year}{2018}).
\newblock \urlprefix\url{https://arxiv.org/abs/1807.08454}.

\bibitem{Fukushima08}
\bibinfo{author}{Fukushima, K.}, \bibinfo{author}{Kahrzeev, D.~E.} \&
  \bibinfo{author}{Warringa, H.~J.}
\newblock \bibinfo{title}{Chiral magnetic effect}.
\newblock \emph{\bibinfo{journal}{Phys. Rev. D}} \textbf{\bibinfo{volume}{78}},
  \bibinfo{pages}{074033} (\bibinfo{year}{2008}).
\newblock \urlprefix\url{https://doi.org/10.1103/PhysRevD.78.074033}.

\bibitem{Liu17}
\bibinfo{author}{Liu, T.}, \bibinfo{author}{Pikulin, D.~I.} \&
  \bibinfo{author}{Franz, M.}
\newblock \bibinfo{title}{Quantum oscillations without magnetic field}.
\newblock \emph{\bibinfo{journal}{Phys. Rev. B}} \textbf{\bibinfo{volume}{95}},
  \bibinfo{pages}{041201(R)} (\bibinfo{year}{2017}).
\newblock \urlprefix\url{https://doi.org/10.1103/PhysRevB.95.041201}.

\bibitem{Lee17}
\bibinfo{author}{Lee, C.~H.} \emph{et~al.}
\newblock \bibinfo{title}{Topolectrical circuits}.
\newblock \emph{\bibinfo{journal}{Communications Physics}}
  \textbf{\bibinfo{volume}{1}}, \bibinfo{pages}{39} (\bibinfo{year}{2018}).
\newblock \urlprefix\url{https://doi.org/10.1038/s42005-018-0035-2}.

\bibitem{Fruchart18b}
\bibinfo{author}{Fruchart, M.} \emph{et~al.}
\newblock \bibinfo{title}{Soft self-assembly of {Weyl} materials for light and
  sound}.
\newblock \emph{\bibinfo{journal}{Proc. Natl. Acad. Sci. USA}}
  \textbf{\bibinfo{volume}{115}}, \bibinfo{pages}{E3655}
  (\bibinfo{year}{2018}).
\newblock \urlprefix\url{https://doi.org/10.1073/pnas.1720828115}.

\bibitem{Soluyanov15}
\bibinfo{author}{Soluyanov, A.~A.} \emph{et~al.}
\newblock \bibinfo{title}{Type-ii weyl semimetals}.
\newblock \emph{\bibinfo{journal}{Nature}} \textbf{\bibinfo{volume}{527}},
  \bibinfo{pages}{495} (\bibinfo{year}{2015}).
\newblock \urlprefix\url{https://dx.doi.org/10.1038/nature15768}.

\end{thebibliography}

\begin{thebibliography}{1}
\expandafter\ifx\csname url\endcsname\relax
  \def\url#1{\texttt{#1}}\fi
\expandafter\ifx\csname urlprefix\endcsname\relax\def\urlprefix{URL }\fi
\providecommand{\bibinfo}[2]{#2}
\providecommand{\eprint}[2][]{\url{#2}}

\bibitem{Xiao15b}
\bibinfo{author}{Xiao, M.}, \bibinfo{author}{Chen, W.-J.}, \bibinfo{author}{He,
  W.-Y.} \& \bibinfo{author}{Chan, C.~T.}
\newblock \bibinfo{title}{Synthetic gauge flux and {Weyl} points in acoustic
  systems}.
\newblock \emph{\bibinfo{journal}{Nature Phys.}} \textbf{\bibinfo{volume}{11}},
  \bibinfo{pages}{920} (\bibinfo{year}{2015}).
\newblock \urlprefix\url{http://dx.doi.org/10.1038/nphys3458}.

\bibitem{Xu17z}
\bibinfo{author}{Xu, Y.}, \bibinfo{author}{Wang, S.-T.} \&
  \bibinfo{author}{Duan, L.-M.}
\newblock \bibinfo{title}{Weyl exceptional rings in a three-dimensional
  dissipative cold atomic gas}.
\newblock \emph{\bibinfo{journal}{Phys. Rev. Lett.}}
  \textbf{\bibinfo{volume}{118}}, \bibinfo{pages}{045701}
  (\bibinfo{year}{2017}).
\newblock
  \urlprefix\url{https://link.aps.org/doi/10.1103/PhysRevLett.118.045701}.

\bibitem{Cerjan18z}
\bibinfo{author}{{Cerjan}, A.}, \bibinfo{author}{{Huang}, S.},
  \bibinfo{author}{{Chen}, K.~P.}, \bibinfo{author}{{Chong}, Y.} \&
  \bibinfo{author}{{Rechtsman}, M.~C.}
\newblock \bibinfo{title}{{Experimental realization of a Weyl exceptional
  ring}}.
\newblock \emph{\bibinfo{journal}{arXiv:1808.09541}}  (\bibinfo{year}{2018}).
\newblock \urlprefix\url{https://arxiv.org/abs/1807.08454}.

\bibitem{Nielsen83}
\bibinfo{author}{Nielsen, H.} \& \bibinfo{author}{Ninomia, M.}
\newblock \bibinfo{title}{The {Adler-Bell-Jackiw} anomaly and {Weyl} fermions
  in a crystal}.
\newblock \emph{\bibinfo{journal}{Phys. Lett. B}}
  \textbf{\bibinfo{volume}{130}}, \bibinfo{pages}{389} (\bibinfo{year}{1983}).
\newblock \urlprefix\url{https://doi.org/10.1016/0370-2693(83)91529-0}.

\end{thebibliography}
	\end{document}